\documentclass[conference]{IEEEtran}
\IEEEoverridecommandlockouts
\usepackage{cite}
\usepackage{amsmath,amssymb,amsfonts}
\usepackage{algorithmic}
\usepackage{textcomp}
\usepackage{xcolor}
\usepackage{caption}
\usepackage{subcaption}
\usepackage{hyperref}

\usepackage{flushend}
\usepackage{fancyhdr}
\usepackage{algorithmic}
\usepackage{graphicx}
\usepackage{textcomp}
\usepackage{xcolor}
\usepackage{tabularx}
\usepackage{colortbl}
\usepackage{tikz}
\usepackage{enumitem}
\usepackage{booktabs}
\usetikzlibrary{arrows, chains, positioning, shapes.geometric, shapes.symbols}
\usetikzlibrary{fadings,calc,patterns}
\usepackage{caption}
\usepackage{subcaption}
\newcolumntype{L}[1]{>{\raggedright\let\newline\\\arraybackslash\hspace{0pt}}m{#1}}
\newcolumntype{C}[1]{>{\centering\let\newline\\\arraybackslash\hspace{0pt}}m{#1}}
\newcolumntype{R}[1]{>{\raggedleft\let\newline\\\arraybackslash\hspace{0pt}}m{#1}}

\begin{document}

\newcommand{\todo}[1]{\textcolor{red}{#1}}

\title{ParaGraph: Weighted Graph Representation for Performance Optimization of HPC Kernels}

\author{\IEEEauthorblockN{Ali TehraniJamsaz\textsuperscript{1}, Alok Mishra\textsuperscript{2}, Akash Dutta\textsuperscript{1}, Abid M. Malik\textsuperscript{3}, Barbara Chapman\textsuperscript{2}, Ali Jannesari\textsuperscript{1}}
\IEEEauthorblockA{\textsuperscript{1}\textit{Iowa State University, Ames, Iowa, USA}}
\IEEEauthorblockA{\{tehrani, adutta, jannesar\}@iastate.edu}
\IEEEauthorblockA{\textsuperscript{2}\textit{Stony Brook University, Stony Brook, New York, USA}}
\IEEEauthorblockA{\{almishra, bchapman\}@cs.stonybrook.edu}
\IEEEauthorblockA{\textsuperscript{3}\textit{Brookhaven National Laboratory, Upton, New York, USA}}
\IEEEauthorblockA{amalik@bnl.gov}
\\[-3.0ex]
}

% \affiliation{%


\maketitle
\begin{abstract}
GPU-based HPC clusters are attracting more scientific application developers
%scientists 
due to their extensive parallelism and energy efficiency.
%Nevertheless, 
% keep abstract short and to the point
%However, due to the increasing variety of GPUs, it is impractical to write separate code that is suitable for each target system for a specific HPC application, considering many factors such as physical memory architectures, offload decisions, data handling, and so on.
%Furthermore, due to the differences in physical memory architectures, an application scientist must ensure data synchronization between the CPU and GPU.
%Other issues with GPU offloading include
%Other challenges involve GPU offloading which include finding appropriate code regions in legacy applications to offload and explicitly handling data between hosts and devices.
% \textcolor{red}{In order to achieve performance portability among a variety of GPU-based systems,} \textcolor{blue}{In order to achieve portability among a variety of multi/many core architectures,} 
In order to achieve portability among a variety of multi/many core architectures, a popular choice for an application developer is to utilize directive-based parallel programming models, such as OpenMP.
However, even with OpenMP, the developer must choose from among many strategies for exploiting a GPU or a CPU. % including the one not using GPU at all.
%Additionally, even if a code is appropriate for a GPU, it might not have enough work to fully exploit a GPU and a CPU execution might be more beneficial.
% As a result, once it has been determined that the code can be offloaded to a GPU, the application developer will greatly benefit from having a tool that will assist them in deciding whether to offload or not, as well as which transformation to use while offloading.
Recently, Machine Learning (ML) approaches have brought significant advances in the optimizations of HPC applications. To this end, several ways have been proposed to represent applications' characteristics for ML models. %However, most of them fail to present the host spot of applications properly, which is crucial to making a good optimization decision.
However, the available techniques fail to capture features that are crucial for exposing parallelism.
In this paper, we introduce a new graph-based program representation for %\textcolor{red}{parallel} \textcolor{blue}{\textit{parallel}} 
\textit{parallel} applications that extends the Abstract Syntax Tree to represent control and data flow information. %which was previously absent in an application representation for ML modeling.
The originality of this work lies in the addition of new 
%\textcolor{blue}{edges exploiting the implicit ordering and parent-child relationships in ASTs, and introducing} 
edges exploiting the implicit ordering and parent-child relationships in ASTs, as well as the introduction of edge weights to account for loop and condition information.
We evaluate our proposed representation by training a Graph Neural Network (GNN) to predict the runtime of an OpenMP code region across CPUs and GPUs.
Various transformations utilizing collapse and data transfer between the CPU and GPU are used to construct the dataset. The predicted runtime of the model is used to determine which transformation provides the best performance.
Results show that our approach is indeed effective and has normalized RMSE as low as $4\times10^{-3}$ to at most $1\times10^{-2}$ in its runtime predictions. 
% We use various transformations utilizing collapse and data transfer between the CPU and GPU as a proof of concept.
% The model determines which transformation provides the best performance by predicting the runtime for each transformation.

%%% Abid %%%%

% 
\end{abstract}


\maketitle

% 20 pages, excluding bibliography
\section{Introduction}
% \textcolor{red}{Over the years, hardware developers have been improving the performance of chips by increasing the number of computing cores. 
% Today, many-core/multi-core processors are ubiquitous; from smartphones to servers, multi-core processing units play a major role in unleashing the computation power.
% One of the communities impacted significantly by many-core technologies is High Performance Computing (HPC).
% To exploit the computation capacity of multi-core processors, applications can be modified by adding Pthreads or OpenMP constructs. }

%\textcolor{blue}{
Over the years, the increase in the number of on-chip cores has led to significant improvement in the performance of parallel code.
To exploit this increased computation capacity, applications have been readily modified using Pthreads or OpenMP constructs. %}
In the last decade, General Purpose Graphic Processing Units (GPGPU) have gained popularity. They can handle massive data parallelism with low power consumption. As a result, most HPC platforms are now coupled with GPGPU(s).
HPC platforms will continue to support more accelerators, but given the difficulties in utilizing and configuring even one accelerator, using and configuring multiple heterogeneous accelerators will become increasingly difficult in the future.
On the other hand, utilizing GPUs effectively imposes challenges that require re-engineering the code and applications. 
The full computing power of GPUs may not be utilized if they are not used properly. In addition, improper data transfer between the host (CPU) and device (GPU) can result in significant overheads.
It is exceptionally challenging and burdensome for developers to create applications for extremely heterogeneous platforms with multiple devices.
The recent emergence of tools and programming models aims to automate this process of application adaptation to heterogeneous platforms. 
One of the most popular parallel programming models, OpenMP~\cite{dagum1998openmp}, aims to make the process of developing parallel programs that can run on different architectures simpler.
Despite this, optimizing a code to use the OpenMP directives correctly is still a tedious task for large and complex applications.

Recent advances in Deep Learning (DL) have enabled researchers to apply DL to a wide range of software engineering jobs and challenges, including code comment generation, compiler optimizations, and heterogeneous platform adaptation.
Since source code cannot be directly fed into DL models, we need a suitable representation for applications to serve as input to various DL models.
In this paper, we propose $ParaGraph$, a graph-based program representation that aims to expose critical characteristics (e.g., the number of iterations in a loop) of HPC applications.
To the best of our knowledge, existing program representations are not designed to expose and represent the characteristics of parallel HPC applications. 
% \textcolor{red}{They usually handle both serial and parallel applications the same way. 
% A machine learning model is prevented from properly inferring knowledge about these types of applications due to the failure to expose the characteristics of parallel applications.}
% \textcolor{blue}{
These program representations usually target serial code and do not incorporate characteristics of a parallel code.
As a result, DL models built on top of these representations cannot model features inherent to parallel codes.%}

%To the best of our knowledge, there are no other works on representing HPC applications for deep learning models. 
% \textcolor{red}{There are a few previous works \cite{mishra2022compoff} that rely on feature engineering to model applications for neural networks. However, feature engineering requires expert knowledge, and it is difficult to adapt it to new trends in developing parallel applications as HPC evolves.
% An adaptable Graph Neural Network (GNN) model is able to automatically learn the characteristics of HPC programs through our proposed representation. 
% Results from experiments show that our model can predict the runtime of an HPC kernel with a very low error rate (at most $3\times10^{-2}$ in terms of normalized RMSE), confirming the efficacy of our strategic approach.}

% \textcolor{blue}{
Previous works, such as \texttt{COMPOFF} \cite{mishra2022compoff}, have relied on feature engineering to overcome these shortcomings.
However, feature engineering requires expert knowledge. 
For a fast-evolving field such as HPC, always relying on expert intervention for such feature engineering is not realistic.
There is a need for an adaptable approach that can automatically extract such features.
Our proposed approach aims to address this and uses Graph Neural Networks (GNNs) to model the code graphs generated by $ParaGraph$.
Experimental results show that our model can predict the runtime of HPC kernels with a very low error rate (at most $1\times10^{-2}$ in terms of normalized RMSE), confirming the efficacy of our strategic approach.
% }

In the rest of the paper, we first discuss some background work in Section~\ref{sec:background} on program representation, OpenMP GPU offloading, DL in compilers, OpenMP Advisor (a tool we used to generate various kernels for our work), and other related work.
Then in Section~\ref{sec:paragraph}, we define $ParaGraph$.
Section~\ref{sec:experiment} covers the experiments carried out in this paper, and Section~\ref{sec:result} discusses the result.
Finally, we conclude our work with discussions of future plans in Section~\ref{sec:conclusion}.
 % 1.5pg
\section{Background and Other Related Work}
\label{sec:background}
In this section, related works and some background for program representation, OpenMP parallelism, and its tools are discussed.

\subsection{Program Representation}
Recently, with breakthroughs in Machine Learning (ML) and specifically Deep Learning (DL), researchers have been applying data-driven approaches to various software engineering tasks, and challenges, ranging from code comment generation \cite{ciurumelea2020suggesting} to compiler optimizations \cite{cummins2021programl}.
However, the source code can not simply be fed to %ML 
DL models as %since ML
DL models do not accept strings as input. 
Therefore, applications need to be presented in a way usable by the models.
Some works have presented programs as a sequence of tokens \cite{raychev2014code, allamanis2016convolutional}. 
While this kind of representation has worked well for some downstream tasks, presenting programs as a sequence of tokens discards critical syntactical and semantic characteristics of programs which are essential for deep learning models to reason. 

Recently, to better present programs' semantics and syntactic information, graph-based representations have been proposed. 
Allamanis et al. \cite{allamanis2017learning} proposed enhanced ASTs and added new edges to present control flow information. 
They applied the augmented AST to two tasks: variable misuse detection and variable name suggestion.
Inspired by their work, we also augment the AST by adding new edges. However, we incorporate additional edges to better represent $loops$ and $if$ conditions.

% \textcolor{red}{On the other hand, at a lower level, Cummins et al. \cite{cummins2021programl} proposed a graph representation using LLVM intermediate representation. 
% This representation aims at solving compiler optimization tasks.}
% \textcolor{blue}{
Cummins et al. in \cite{cummins2021programl} proposed a lower-level graph representation based on LLVM intermediate representation for solving compiler optimization tasks.%}
%While 
These representations are effective for downstream tasks, such as simple algorithm classification. To the best of our knowledge, there does not exist a representation tailored toward representing the characteristics of \textit{parallel} loops and if-statements. 
% Therefore, current existing representations fail to present HPC kernels properly.

\subsection{OpenMP GPU Offloading}
When different architecture supports either a different native language or various optimization techniques for the same language, program portability becomes a serious problem. 
Regrettably, programming languages and compilers are still far from being able to handle portability on their own. 
Program portability should be the primary priority for developers who have to deal with multiple device memory accesses on the same data or someone whose programs must work effectively on systems with various node architectures, such as many-core vs. GPU-accelerated nodes.

Utilizing a directive-based programming paradigm, such as OpenMP, the de-facto standard for parallel programming in C/C++ and Fortran, is one approach to ensuring portability across several architectures.
OpenMP intends to move to extremely heterogeneous architectures~\cite{heroux2020ecp} and has supported GPU offloading from specification 4.0.
Unfortunately, even with OpenMP, optimizing large-scale applications remains a challenging task.

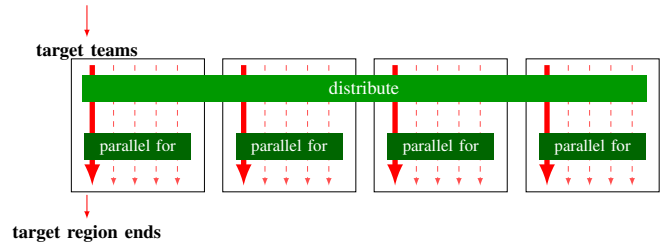
\begin{figure}[b!]
\vspace{-1em}
\centering
\resizebox {\columnwidth} {!} {
\begin{tikzpicture} 

\node[draw, minimum width=2.5cm, minimum height=2.5cm, anchor=south west] at (0.5cm, 0.5cm)(team1){};
\node[draw, minimum width=2.5cm, minimum height=2.5cm, anchor=south west, right=0.33cm of team1](team2) {};
\node[draw, minimum width=2.5cm, minimum height=2.5cm, anchor=south west, right=0.33cm of team2](team3) {};
\node[draw, minimum width=2.5cm, minimum height=2.5cm, anchor=south west, right=0.34cm of team3](team4) {};

\foreach \i in {1,...,4} {
    \foreach \j in {1,...,5} {
        \node[circle, fill=none, minimum size=1pt, inner sep=0pt] (top\i\j) at ($(team\i.north west)+(0.4*\j,-0.1cm)$) {};
        \node[circle, fill=none, minimum size=1pt, inner sep=0pt] (bottom\i\j) at ($(team\i.south west)+(0.4*\j,0.1cm)$) {};
    }
    \foreach \j in {2,...,5} {
        \draw[-latex, dashed, thin, red!70] (top\i\j)--(bottom\i\j);
    }
    \draw[-latex, line width=1mm, red] (top\i1)--(bottom\i1);
    \node[fill=green!40!black, minimum width=2cm, minimum height=0.5cm, anchor=west, below=1.4cm of team\i.north, align=center, inner sep=1](dist) {\textcolor{white}{\small{parallel for}}};
}

\node[fill=green!60!black, minimum width=10.6cm, minimum height=0.5cm, anchor=west, below right=0.3cm and 0.2cm of team1.north west, align=center](dist) {\textcolor{white}{distribute}};

\draw[-latex,red] ($(team1.north west)+(0.3cm, 1cm)$)--($(team1.north west)+(0.3cm, 0.4cm)$) node[below=0] {\textcolor{black}{\textbf{target teams}}};
\draw[-latex,red] ($(team1.south west)+(0.3cm, -0.05cm)$)--($(team1.south west)+(0.3cm, -0.5cm)$) node[below=0] {\textcolor{black}{\textbf{target region ends}}};

\end{tikzpicture}
}
\caption{OpenMP target teams distribute parallel for.}
\label{fig:teams}
\vspace{-0.5em}
\end{figure}

Since GPUs are very parallel machines, programmers should take full advantage of this fact to get the most performance possible.
The ``\texttt{omp parallel for}'' pragma alone will only parallelize a code for CPUs; it won't offload the computation to a GPU.
We expect a high level of coarse grain parallelism on a platform like GPU.
The amount of parallelism that a GPU can use is constrained by this design.
Figure~\ref{fig:teams} illustrates how OpenMP \texttt{teams} and \texttt{distribute} directives create additional levels of parallelism.
At the start of a target region, only one team and one member thread are active. 
The \texttt{teams distribute} directive distributes the full loop iteration space among all available teams. 
Additionally, in the case of nested parallel loops, we can utilize the \texttt{parallel for} directive on them to further distribute the iterations of the nested loops among threads within a team.
We utilize the combined directive ``\texttt{teams distribute parallel for}'' to distribute the iteration space of one loop among teams and threads inside a team when there is just one level of parallel loops or when the outer loop has sufficient parallelism.
\texttt{teams} are used to group threads, and \texttt{distribute} enables a team group to be scheduled to run in a loop. 
Teams resemble CUDA threadblocks on NVIDIA GPUs, as there is no synchronization primitive to function as a barrier between threads from different teams.
And the threads within each team are parallelized typically using \texttt{parallel for}.

There are several frameworks being worked on right now that will automatically assist application developers in handling severe heterogeneity.
These frameworks necessitate analytical cost models, which will help them select the kind of optimization the application requires.
Hand-tuned cost functions are extensively employed currently; however, calculating optimization costs requires a deeper understanding of the underlying hardware.
Despite its efficacy, manually building a cost model for a single architecture can take months.
Since cost functions are crucial and manual tuning is time-consuming, compiler engineers are investigating Machine, and Deep Learning approaches as a means of automating this process.

\subsection{ML in Compiler}
Recently, a lot of research has been done on how to use learning-based techniques in compilation as well.
Early work exploiting ML in compilers primarily explored its use to help optimize sequential programs.
However, its application to the task of optimizing parallel programs has attracted significant attention in the past decade due to the prevalence of multi-core platforms and, more recently, heterogeneous systems~\cite{kim2020comprehensive,rzadca2020autopilot}. 
Mirka \textit{et.al.}~\cite{mirka2020online} devised a decision-tree-based approach to predict the scheduling policy for an OpenMP parallel region.
Also, Denoyelle \textit{et.al.}~\cite{denoyelle2019data} uses machine learning techniques to optimize OpenMP programs for scheduling policies and the number of threads.
Alcaraz \textit{et.al.}~\cite{alcaraz2022predicting} uses neural networks and performance counters to predict the number of threads. Dutta \textit{et.al.}~\cite{dutta2022pattern} uses an LLVM-IR based graph representation and performance counter to train a GNN model to predict number of thread, chunk size and scheduling policy for OpenMP loops. 
Tree and graph-based features have also been used by Malik \textit{et.al.}~\cite{malik2011automatic}, who present a unique graph-based approach for feature representation.
Learning-based techniques were used to build classifiers to determine whether to offload OpenCL code~\cite{dublish2019poise} and to select a clock frequency at which the processor should operate~\cite{iranfar2019machine}. 
A high level of accuracy was reported; however, the benefits could not be quantified as the work did not attempt to generate modified code.
They also explored regression techniques to build curve fitting models to search for the sweet spot for work partitioning among processors~\cite{grewe2011static} or a trade-off of energy and performance~\cite{sayadi2018energy}.

The results of prior efforts from applying ML on compiler optimizations are encouraging. 
However, new feature engineering practices need to be explored that can help learning-based methods to learn more about a code and its computational needs. 
In \texttt{COMPOFF}~\cite{mishra2022compoff}, the authors provide a proof of concept for using an Artificial Neural Network model to make better decisions in offloading a kernel to a GPU using OpenMP.
They utilize a fully-connected feed-forward network, also referred to as multi-layer perceptrons (MLPs), which are effectively stacked layers of linear regression.
% However, their model necessitates calculating how many operations are included within a kernel, which is a difficult task in and of itself.

\subsection{OpenMP Advisor}
\label{sec:ompadvisor}
OpenMP Advisor~\cite{mishra2023openmp} is a first-of-its-kind compiler tool that enables OpenMP code offloading to a GPU via Machine Learning.
The Advisor is divided into three major modules: Kernel Analysis, Cost Model, and Code Transformation.
The Kernel Analysis module recommends various variants for a given application, and the Code Transformation module generates codes for those variants.
In our work, we use the code transformation module of OpenMP Advisor to generate various kernels for training our model.
Through the use of \texttt{COMPOFF}, a machine learning-based cost model, this tool can identify the kernels that are most suitable for offloading by predicting their runtime.
But \texttt{COMPOFF} has some limitations.
It requires figuring out how many operations are contained within a kernel, which is a challenging task in and of itself.
The Advisor's current functionality is only restricted to GPUs, though it has the potential to be expanded to support additional OpenMP-capable hardware.
One technique for expanding to other devices is to develop a model that works outside GPUs as well.
Extending \texttt{COMPOFF} to work beyond GPUs is a daunting task, and there is a need for a new cost model that can bridge this gap.

\subsection{Other Related Works}
A recent tool called BLISS~\cite{roy2021bliss} allows for auto-tuning parallel applications without the use of instrumentation or domain-specific knowledge.
This auto-tuner develops a variety of lightweight models with the aid of Bayesian optimization that can compete with the output quality of a complex, heavyweight Machine Learning (ML) model.
In the same context, Bayesian optimization is frequently used by other autotuners, including BOHB~\cite{falkner2018bohb}, HiPerBOt~\cite{menon2020auto}, GPTune~\cite{liu2021gptune} and ytopt~\cite{wu2022autotuning}.
Unfortunately, due to their expensive evaluation functions, tuning large-scale HPC systems, which are becoming increasingly heterogeneous, complex, and expensive to evaluate, is still quite challenging.
Nonetheless, most of the autotuners mentioned above are online, meaning that even in the inference time, they execute applications iteratively to tune them. Therefore, in terms of complex heterogeneous large-scale HPC systems, these tuners will be very costly to apply. In contrast, $ParaGraph$ is an offline approach and does not require the execution of applications in the inference phase.
% These tools are unsuitable for our work due to their insufficient support for heterogeneous architectures, such as GPUs.
\texttt{COMPOFF} ~\cite{mishra2022compoff} provides a novel portable cost model that statically calculates the cost of OpenMP offloading on various architectures. COMPOFF is also an offline approach. % using a neural network model.
 In this work, we compare $ParaGraph$ to \texttt{COMPOFF} since both can predict the runtime of a kernel on a GPU statically. % 1 pg
\section{ParaGraph}
\label{sec:paragraph}
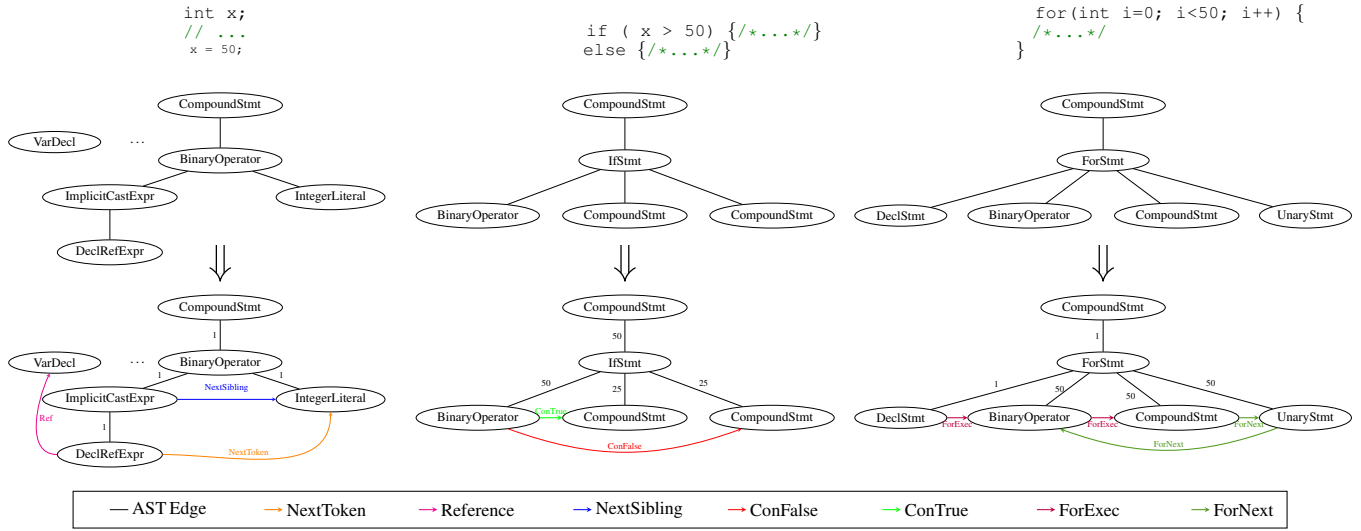
\begin{figure*}[hbt!]
\centering
\tikzset{
empty/.style = {ellipse, minimum width=2.5cm},
base/.style = {draw, ellipse, minimum width=2.5cm},
forexec/.style = {->, >=stealth, thick, purple},
fornext/.style = {->, >=stealth, thick, purple!40!green},
legend/.style = {minimum width=2.5cm, anchor=west, text width=2.3cm},
}
\resizebox{\textwidth}{!} {
\scriptsize
\begin{tikzpicture}
    \node[empty,anchor=west] at (3,14) {\Large \texttt{int x;}};
    \node[empty,anchor=west] at (3,13.5) {\Large \texttt{\textcolor{green!50!black}{// ...}}};
    \node[empty,anchor=west] at (3.15,13) {\normalsize \texttt{x = 50;}}; 
    \node[base](csbo1) at (4.5,11.5) {\normalsize CompoundStmt};
    \node[base](bobo1) at (4.5,10) {\normalsize BinaryOperator};
    \node[base](icebo1) at (1.5,9) {\normalsize ImplicitCastExpr};
    \node[base](inbo1) at (7.5,9) {\normalsize IntegerLiteral};
    \node[base](varbo1) at (0,10.5) {\normalsize VarDecl};
    \node[empty] at (2.25,10.5) {\textbf{. . .}};
    \node[base](drebo1) at (1.5,7.5) {\normalsize DeclRefExpr};
    
    \draw[-] (csbo1) -- (bobo1);
    \draw[-] (bobo1) -- (icebo1);
    \draw[-] (bobo1) -- (inbo1);
    \draw[-] (icebo1) -- (drebo1);
	
	\draw[-implies,line width=1pt,double distance=5pt] (4.5,7.7) -- (4.5,6.7);
	
    \node[base](csbo1) at (4.5,6) {\normalsize CompoundStmt};
    \node[base](bobo1) at (4.5,4.5) {\normalsize BinaryOperator};
    \node[base](icebo1) at (1.5,3.5) {\normalsize ImplicitCastExpr};
    \node[base](inbo1) at (7.5,3.5) {\normalsize IntegerLiteral};
    \node[base](varbo1) at (0,4.5) {\normalsize VarDecl};
    \node[empty] at (2.25,4.5) {\textbf{. . .}};
    \node[base](drebo1) at (1.5,2) {\normalsize DeclRefExpr};
    
    \draw[-] (csbo1) -- node[pos=0.5, left] {1} (bobo1);
    \draw[-] (bobo1) -- node[pos=0.5, above left] {1} (icebo1);
    \draw[-] (bobo1) -- node[pos=0.5, above right] {1} (inbo1);
    \draw[-] (icebo1) -- node[pos=0.5, left] {1} (drebo1);
	\draw [->,>=stealth,orange] (drebo1) to[out=0,in=-90] node[pos=0.5, above left] {NextToken}(inbo1);
	\draw [->,>=stealth,blue] (icebo1) -- node[pos=0.5, above=0.1] {NextSibling}(inbo1);
	\draw [->,>=stealth,magenta] (drebo1) to[out=180,in=-115] node[pos=0.5, above right] {Ref}(varbo1);
	
%%%%%%%%%%%%

    \node[empty,anchor=west] at (13,13.5) {\Large \texttt{if ( x > 50) \{\textcolor{green!50!black}{/*...*/}\}}};
    \node[empty,anchor=west] at (13.4,13) {\Large \texttt{else \{\textcolor{green!50!black}{/*...*/}\}}};
    
	\node[base](csif1) at (15.5,11.5) {\normalsize CompoundStmt};
	\node[base](if) at (15.5,10) {\normalsize IfStmt};
	\node[base](bo) at (11.5,8.5) {\normalsize BinaryOperator};
	\node[base](csif2) at (15.5,8.5) {\normalsize CompoundStmt};
	\node[base](csif3) at (19.5,8.5) {\normalsize CompoundStmt};
	
	\draw[-] (csif1) -- (if);
	\draw[-] (if) -- (bo);
	\draw[-] (if) -- (csif2);
	\draw[-] (if) -- (csif3);
	
	\draw[-implies,line width=1pt,double distance=5pt] (15.5,7.7) -- (15.5,6.7);
	
	\node[base](csif1) at (15.5,6) {\normalsize CompoundStmt};
	\node[base](if) at (15.5,4.5) {\normalsize IfStmt};
	\node[base](bo) at (11.5,3) {\normalsize BinaryOperator};
	\node[base](csif2) at (15.5,3) {\normalsize CompoundStmt};
	\node[base](csif3) at (19.5,3) {\normalsize CompoundStmt};
	
	\draw[-] (csif1) -- node[pos=0.5, left] {50} (if);
	\draw[-] (if) -- node[pos=0.5, above left] {50} (bo);
	\draw[-] (if) -- node[pos=0.5, left] {25} (csif2);
	\draw[-] (if) -- node[pos=0.5, above right] {25} (csif3);
	\draw[->,>=stealth,thick,green] (bo) -- node[pos=0.5, above] {ConTrue} (csif2);
	\draw [->,>=stealth,red] (bo) to[out=-20,in=-160] node[pos=0.5, above] {ConFalse}(csif3);

 %%%%%%%%%%%%

    \node[empty,anchor=west] at (25,14) {\Large \texttt{for(int i=0; i<50; i++) \{ }};
    \node[empty,anchor=west] at (26,13.5) {\Large \texttt{\textcolor{green!50!black}{/*...*/}}};
    \node[empty,anchor=west] at (25,13) {\Large \texttt{\}}};
    
	\node[base](csfor1) at (28.5,11.5) {\normalsize CompoundStmt};
	\node[base](for) at (28.5,10) {\normalsize ForStmt};
	\node[base](dsfor) at (23,8.5) {\normalsize DeclStmt};
	\node[base](bofor) at (26.5,8.5) {\normalsize BinaryOperator};
	\node[base](csfor2) at (30.5,8.5) {\normalsize CompoundStmt};
	\node[base](usfor) at (34,8.5) {\normalsize UnaryStmt};
	
	\draw[-] (csfor1) -- (for);
	\draw[-] (for) -- (dsfor);
	\draw[-] (for) -- (bofor);
	\draw[-] (for) -- (usfor);
	\draw[-] (for) -- (csfor2);
	
	\draw[-implies,line width=1pt,double distance=5pt] (28.5,7.7) -- (28.5,6.7);
	
	\node[base](csfor1) at (28.5,6) {\normalsize CompoundStmt};
	\node[base](for) at (28.5,4.5) {\normalsize ForStmt};
	\node[base](dsfor) at (23,3) {\normalsize DeclStmt};
	\node[base](bofor) at (26.5,3) {\normalsize BinaryOperator};
	\node[base](csfor2) at (30.5,3) {\normalsize CompoundStmt};
	\node[base](usfor) at (34,3) {\normalsize UnaryStmt};
	
	\draw[-] (csfor1) -- node[pos=0.5, left] {1} (for);
	\draw[-] (for) -- node[pos=0.5, above left] {1} (dsfor);
	\draw[-] (for) -- node[pos=0.5, left] {50} (bofor);
	\draw[-] (for) -- node[pos=0.5, above right] {50} (usfor);
	\draw[-] (for) -- node[pos=0.5, below left] {50} (csfor2);
	\draw [forexec] (dsfor) -- node[pos=0.5, below] {ForExec} (bofor);
	\draw [forexec] (bofor) -- node[pos=0.5, below] {ForExec} (csfor2);
	\draw [fornext] (csfor2) -- node[pos=0.5, below] {ForNext} (usfor);
	\draw [fornext] (usfor) to[out=-160,in=-20] node[pos=0.5, above] {ForNext}(bofor);
	
	\draw[draw=black,thick] (0.5,1) rectangle (33.5,0);
	\node[legend](ast) at (2,0.5) {\Large AST Edge};
	\node[legend](nxtTok) at (6.2,0.5) {\Large NextToken};
	\node[legend](ref) at (10.4,0.5) {\Large Reference};
	\node[legend](nxtSib) at (14.6,0.5) {\Large NextSibling};
	\node[legend](conF) at (18.8,0.5) {\Large ConFalse};
	\node[legend](conT) at (23,0.5) {\Large ConTrue};
	\node[legend](forEx) at (27.2,0.5) {\Large ForExec};
	\node[legend](forNxt) at (31.4,0.5) {\Large ForNext};
	
	\draw[-] (ast.west) -- ($(ast.west)-(.5,0)$);
	\draw[<-,>=stealth,thick, orange] (nxtTok.west) -- ($(nxtTok.west)-(.5,0)$);
	\draw[<-,>=stealth,thick,blue] (nxtSib.west) -- ($(nxtSib.west)-(.5,0)$);
	\draw[<-,>=stealth,thick,magenta] (ref.west) -- ($(ref.west)-(.5,0)$);
	\draw[<-,>=stealth,thick,red] (conF.west) -- ($(conF.west)-(.5,0)$);
	\draw[<-,>=stealth,thick,green] (conT.west) -- ($(conT.west)-(.5,0)$);
	\draw[<-,>=stealth,thick,purple] (forEx.west) -- ($(forEx.west)-(.5,0)$);
	\draw[<-,>=stealth,thick,purple!40!green] (forNxt.west) -- ($(forNxt.west)-(.5,0)$);
\end{tikzpicture}
}
\caption{Modification to AST to create Augmented AST for ParaGraph}

\label{augmentedAST}
\end{figure*}

In this section, we present $ParaGraph$, a weighted graph representation of programs that captures characteristics related to HPC kernels. 
Abstract Syntax Tree serves as a base for $ParaGraph$.
In fact, additional edges and attributes are introduced to AST to construct $ParaGraph$.
Some previous works have proposed various ways of augmenting ASTs to present semantic information \cite{allamanis2022graph, allamanis2017learning}.
Although these AST-based program representations are effective in training deep learning models for downstream tasks such as variable misuse detection, not enough attention has been given to better represent regions of programs that have a high impact on the performance of the application, such as $loops$ and $if$ condition. In $ParaGraph$ we try to address these shortcomings.
%to the best of our knowledge there does not exist any work to target \texttt{loops} and \texttt{if} conditions.
As stated, $ParaGraph$ is based on AST; however, it includes new edges to present characteristics such as control flow or order of children of a node. Moreover, $loops$ are represented by special edges conveying the order in which the loop condition and its body execute iteratively. 
% As mentioned, ParaGraph is built on top of AST and includes additional attributes to not only preserve semantic and syntactical information but also present the number of iterations in loops, if statements in a more elegant way by adding weights to the edges.
Additionally, we add weights to the edges to present how many times each node will be utilized during the execution of a program.
% Moreover, edge weights include implicit information about how workload is distributed among threads.
It is worth noting that all these augmentations are applied statically; therefore, they do not cause significant overhead.
$ParaGraph$ can be directly utilized by Graph Neural Network models, and our experimental results show that, in performance optimization, this representation is quite effective and outperforms the state-of-the-art approach. 

% Define the graph
% How graph is construction
\subsection{ParaGraph Construction}
\subsubsection{Abstract Syntax Tree} As stated, the base of $ParaGraph$ is Abstract Syntax Tree.
Typically, any kind of compiler can be used to generate AST.
In this study, we use \texttt{Clang}, a popular \texttt{C/C++} front-end, to parse and compile \texttt{C/C++} programs.
ASTs contain two types of nodes: \textit{non-terminal} and \textit{terminal}.
Non-terminal nodes are often called \textit{syntax nodes}, whereas terminal nodes are called \textit{syntax tokens}.
The relation between nodes in AST is a parent-child relation.
Once an AST is retrieved, we add additional edges and attributes depicting information related to control flow and data flow. 
The following subsections will provide more details on these augmentations.

\subsubsection{Augmenting Abstract Syntax Tree}
\label{sec:augment_ast}
ASTs typically provide structural and syntactical information of a program. 
Although this information can be useful for neural networks to learn the characteristics of programs to some extent, it has been shown that adding additional attributes and information, especially from a compiler point of view, boosts the learning capabilities of deep learning models.
Inspired by existing works \cite{allamanis2022graph, allamanis2017learning}, we introduce the following additional edges to AST (shown in Figure \ref{augmentedAST}) to represent information related to the control and data flow of the programs.
\begin{itemize}[nosep,leftmargin=*]
    \item[--] \textbf{NextToken: } By default there is no order imposed among the \textit{syntax tokens}. 
    However, from a compiler's perspective, syntax tokens have an order. 
    The order among the syntax tokens is from far left token to the far right one.
    To present this information in a graph, we introduce a new edge type called \texttt{NextToken}. 
    \texttt{NextToken} connects each syntax token to the syntax token on its right side. In other words, it connects each \textit{syntax token} to the next \textit{syntax token}. In this way, an order is imposed on syntax tokens. This edge type is shown in orange in Figure \ref{augmentedAST} (the AST on the left).
    
    \item[--] \textbf{NextSib: } As mentioned above, edges in ASTs show a parent-child relationship. 
    Compilers assume an order among children nodes. 
    For example, a binary operator such as division has two children. In an AST, the left child is always the numerator, and the right child is the denominator. 
    Therefore, it is necessary to show the order of the children. \texttt{NextSib} presents this information. 
    \texttt{NextSib} connects each syntax node to its sibling on the right (the blue edge in Figure \ref{augmentedAST}.
    
    \item[--] \textbf{Ref: } In clang's AST, referenced variables are presented by \texttt{DecRefExpr} nodes. 
    These nodes are terminal and do not have children. 
    During graph construction, we introduce \texttt{Ref} edges (shown in pink in Figure \ref{augmentedAST}) connecting a \texttt{DecRefExpr} node to where the corresponding variable is defined.
    This conveys information where a declared variable is used throughout the graph. 
    % In order to present, where a declared variable is used throughout the code, we introduce \texttt{Ref} edges, that connects a \texttt{DecRefExpr} to where the corresponding variable is defined.
    
    \item[--] \textbf{ForNext, ForExec: } Loops are typically shown as \texttt{ForStmt} nodes in Clang's AST.
    \texttt{ForStmt} nodes usually have four children.
    First child is related to initializing the loop counter, whereas the second child is about checking the loop's condition, deciding whether the next iteration should be executed or the loop has ended. 
    Third child is \texttt{CompoundStmt} which presents the body of a loop.
    Lastly, the last child is a modifier that modifies the loop counter such as incrementing it by one.
    The relationship between these four children exposes critical characteristics of loops. 
    These characteristics are known by compilers but are not presented explicitly in the ASTs.
    We add \texttt{ForExec} edges, which connect the first child (counter initialization) to the second child (loop condition) and also connect the second child to the third child (loop body) as shown in the right side of Figure \ref{augmentedAST}. 
    In fact, \texttt{ForExec} edges show the flow of executing the next iteration of the loop. 
    We connect the third child (Loop's body) of the \texttt{ForStmt} node to the fourth child (loop counter modifier) via a new type of edge called \texttt{ForNext}. 
    Basically, \texttt{ForNext} represents the information related to whether the next iteration of the loop needs to be executed or not, whereas \texttt{ForExec} represents the next execution of the loop.  
    After the fourth child (loop counter modifier) comes the second child (loop condition), which checks if next iteration is going to happen or if the loop has ended. Therefore, the fourth child is connected to the second child with a \textit{ForNext} edge as shown in Figure~\ref{augmentedAST}.
    
    \item[--] \textbf{ConTrue, ConFalse: } If statements in the ASTs are shown by \texttt{IfStmt} nodes.
    \texttt{IfStmt} nodes typically have three children.
    First child presents the comparison condition; the second child is the body of the \texttt{if} condition, and the third child is the body of the \texttt{else} part.
    To present this information, we connect the first child to the second child through a new type of edge called \texttt{ConTrue} to show that the flow moves to the second child if the \texttt{if statement} is true.
    On the other hand, the first child is connected to the third child via a \texttt{ConFalse} edge.
    This edge shows that if the condition of \texttt{if} statement is not met, the flow moves to the third child.
    
    \item[--] \textbf{Child:} Child edges are normal AST edges that present a parent-child relationship between the nodes in AST.
    
\end{itemize}

% include a flow diagram
\begin{figure*}[ht!]
    \usetikzlibrary{shapes.symbols,shapes.geometric,shadows}
\usetikzlibrary{shapes.arrows}
\tikzset{
base/.style = {fill, circle},
runtime/.style = {draw, minimum width=1.8cm, minimum height=0.3cm},
bl/.style = {->, >=stealth, line width=0.25mm, blue!50},
pl/.style = {->, >=stealth, line width=0.25mm, purple!50},
yl/.style = {->, >=stealth, line width=0.25mm, orange!50},
myarrow/.style = {single arrow, draw, minimum width = 10pt, single arrow head extend=2pt, minimum height=6mm},
multidocument/.style={
    shape=rectangle,
    draw,
    fill=white,
    tape bend top=1,
    double copy shadow={shadow xshift=-0.5ex, shadow yshift=0.5ex}},
}

\sbox 0{
\begin{tikzpicture}[scale=0.350]
    \node[base](n) {}
        child {node[base] {}
            child{node[base]{}}
        }
        child {node[base]{}
            child[missing]
            child{node[base] {}
                child{node[base]{}}
                child{node[base]{}}
            }
            child{node[base]{}}
        };
    \node[below=1.5 of n] (ast) {\normalsize{AST}};
        
\end{tikzpicture}
}
\sbox 1{
\begin{tikzpicture}[scale=0.350,font=\tiny]
    \node[base] (n1) {}
        child {node[base](n2){}
            child{
                node[base](n3){}
                edge from parent node[left]{1}
            }
            edge from parent node[left]{1}
        }
        child {node[base](n4){}
            child[missing]
            child{node[base] (n5) {}
                child{
                    node[base](n6){}
                    edge from parent node[left=-0.1]{N}
                }
                child{
                    node[base](n7){}
                    edge from parent node[right=-0.1]{N}
                }
                edge from parent node[left]{1}
            }
            child{
                node[base](n8){}
                edge from parent node[right]{1}
            }
            edge from parent node[right]{1}
        };
        \draw[yl] (n3) -- (n6);
        \draw[yl] (n6) -- (n7);
        \draw[yl] (n7) -- (n8);
        \draw[bl] (n2) -- (n4);
        \draw[bl] (n5) -- (n8);
        \draw[pl] (n6) to[out=180,in=180] (n2);
        
    \node[below=1.5 of n1] (para) {\normalsize{ParaGraph}};
\end{tikzpicture}
}

\sbox 2{
\begin{tikzpicture}[scale=0.5]
    \node[multidocument, minimum width=0.8cm, minimum height=1cm](var1) {}; 
    \node[multidocument, minimum width=0.8cm, minimum height=1cm, below=0.25 of var1](var2) {}; 
    \node[multidocument, minimum width=0.8cm, minimum height=1cm, below=0.25 of var2](var3) {}; 
    \node[multidocument, minimum width=0.8cm, minimum height=1cm, below=0.25 of var3](var4) {}; 
    \node[below=0 of var4] {\small{Variants}};
\end{tikzpicture}
}

\sbox 3{
\begin{tikzpicture}[scale=0.5]
    \node[runtime](head) {\normalsize\textbf{Time (ms)}};
    \node[runtime, below=-0.017 of head](row1) {\small 746846};
    \node[runtime, below=-0.017 of row1](row2) {\small 9984695};
    \node[runtime, below=-0.017 of row2](row3) {\small 916548};
\end{tikzpicture}
}

\resizebox{\textwidth}{!} 
{
\begin{tikzpicture}
    \node[multidocument, minimum width=0.8cm, minimum height=1cm](doc) {\tiny\textit{Input}}; 
    \node[draw, rectangle, rounded corners=10, minimum width=2cm, minimum height=1.5cm, text width=1.8cm, align=center, below=0.5 of doc] (gen) {\large{Variant Generator}};
    
    \node[right=0.5 of gen] (variants) {\usebox 2} ;
    \node[draw, rectangle, rounded corners=10, minimum width=2cm, minimum height=1.5cm, text width=1.8cm, align=center, below right=0.5 and 0.75 of variants.north east, anchor=north west] (compiler) {\large{Compiler Frontend}};
    \node[right=0.5 of compiler] (ast) {\usebox 0} ;
    \node[draw, rectangle, rounded corners=10, minimum width=2cm, minimum height=1.5cm, text width=1.8cm, align=center, right=0.5 of ast] (para) {\large{ParaGraph Generator}};
    \node[right = 0.5 of para](graph) {\usebox 1};
    
    \path ($(variants.north east)+(0.5,0.25)$) -- ($(graph.north east)+(0.25,0.25)$) node[midway](paraText) {\normalsize\textbf{ParaGraph Generation}};

    \node[draw, rectangle, rounded corners=10, minimum width=2cm, minimum height=1.5cm, text width=1.8cm, align=center, above right=0.4 and 1 of variants.south east, anchor=south west] (compile) {\large{Compile and Build}};
    \node[draw, rectangle, rounded corners=10, minimum width=3.5cm, minimum height=1.5cm, text width=3.5cm, align=center, right=0.75 of compile] (execute) {\large{ Runtime Measurement Module}};
    \node[right = 0.5 of execute](runtime) {\usebox 3};
    
    \node[below=2.9 of paraText] {\normalsize\textbf{Measuring Execution Time of Applications}};

    \draw[dotted,thick, red] ($(variants.north east) + (0.5, 0)$) rectangle ($(graph.south east) +(0.25,0)$);
    \draw[dotted,thick, red] ($(variants.south east) + (0.5, 0)$) rectangle ($(graph.south east) +(0.25,-0.15)$);
    \draw[->, thick] (doc) -- (gen);
    \draw[->, thick] (gen) -- ($(variants.west)+(0.2,0)$);
    \draw[->, thick] ($(variants.east)-(0.3,0)$) -- ($(variants.east)+(0.1,0)$) |- (compiler);
    \draw[->, thick] ($(variants.east)+(0.1,0)$) |- (compile);
    
    \draw[->, thick] (compiler) -- ($(ast.west)+(0.15,0)$);
    \draw[->, thick] ($(ast.east)-(0.1,0)$) -- (para);
    \draw[->, thick] (para) -- ($(graph.west)+(0.25,0)$);
    
    \draw[->, thick] (compile) -- (execute);
    \draw[->, thick] (execute) -- ($(runtime.west)+(0.2,0)$);
    
    \node[draw, cylinder, shape border rotate = 90, shape aspect=.25, minimum height=1.3cm, align=center, right=1.1 of runtime] (dataset) {\large{Dataset}};
    
    \node[draw,circle, below right=0.3 and 1.25 of graph.east, text width=1cm, align=center](model) {\large{GNN model}};
    \node[draw, above=0.4 of model, minimum width=1.2cm, minimum height=1cm, align=center] (prediction) {\tiny\textit{Prediction}};
    
    \draw[->,thick] ($(graph.east)-(0.2,0)$) -- node[above, pos=0.75]{\tiny{Predict}} ($(graph.east)+(0.8,0)$) |- (model);
    
    \draw[->,thick] ($(graph.south)+(0,0.2)$) -- (graph.south) -- ($(graph.south)+(1.5,0)$) node[below right=-0.1,text width=2mm, align=center]{\tiny{\shortstack{T\\r\\a\\i\\n}}} |-  (dataset);
    \draw[->,thick] ($(runtime.east)-(0.2,0)$) -- (dataset);
    
    \draw[->,thick] (dataset) -- node[right, pos=0.5]{\tiny{Train}} (model);
    \draw[->,thick] (model) -- (prediction);
    
\end{tikzpicture}
}
	\caption{Workflow of ParaGraph}
	\label{fig:flow}
    \vspace{-1em}
\end{figure*}
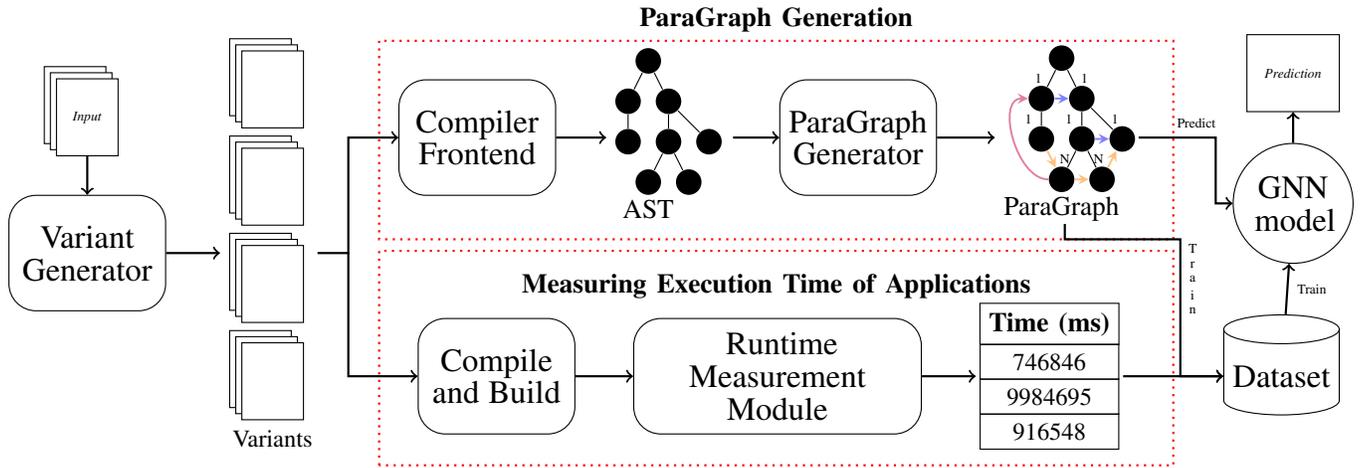

\subsubsection{Weighted Edges}
In the previous section, we explained how we augment the AST by introducing new types of edges to better present programs.
However, there is still some information that is missing in our representation.
For instance, whenever we encounter a loop in a program, the loop's body could be executed multiple times.
Or the branches of an if statement will not execute the same number of times.
To present this information in our graph structure, we augment the edges by adding weights.
Weights of edges are calculated based on the region and edge type.
Weights are added only to \texttt{Child} edges since these are AST edges, and a compiler uses the AST nodes and child edges to transform an AST to lower-level for machine execution. For illustration purpose, some edge weights are shown in Figure \ref{augmentedAST}.
We add the weights as follows:
\begin{itemize}[nosep,leftmargin=*]
    \item[--] \textbf{Default edge weight:} By default, we add a weight of 1 to edges to represent the fact that each statement in code executes one time only, and once a statement executes, the control moves to the next statement.
    \item[--] \textbf{Loops:} Despite the typical statements that execute only once, statements in the body of a loop will be executed several times depending on the number of iterations. 
    To expose the number of iterations in our graph representation, we first observe the number of iterations in a loop and then multiply the edge weights by that number. 
    Moreover, the workload of each thread is also implicitly taken into account if the loop scheduling is static. 
    This is done by dividing the number of iterations by the number of threads.
    For instance, if a loop has 100 iterations, and it is statically scheduled among four threads, we roughly assume each thread executes 25 iterations; therefore, edges within the body of the loop will have a weight of 25.
    \item[--] \textbf{If statements:} The nodes and edges under one branch of an if statement versus the other will not have the same number of executions. 
    Therefore, there is a need for justification of edges of \texttt{if} statements.
    To illustrate the execution of branches, we posit a probability. The assumption is that each branch of an \texttt{if} statement has a probability of $\frac{1}{2}$. 
    Therefore, the weights of edges within if statements are divided by 2. 
    Experimental results show that this representation of \texttt{if} statements provides better results, further optimization of edge weights is left as a future work. 
    %For example, it might be possible to use techniques to estimate the probability of each branch more precisely.
    
\end{itemize}

% How it benefits GNN
\subsection{Benefit to GNN} 
$ParaGraph$ can be used by existing GNN models, which are a type of neural network that can operate over graphs, for downstream tasks. Especially, those tasks that are related to performance optimizations.
$ParaGraph$ can be exploited both as a heterogeneous graph or as a homogeneous graph.
For simplicity, in this study, we treat $ParaGraph$ as a homogeneous graph.
Typically a graph is shown as Equation~\ref{simple_graph_definition}, where $V$ is a set of nodes and $E$ contains an adjacency matrix. 
Elements of $E$ are either 1 or 0 to show whether an edge exists between two nodes.

\begin{equation}
    G = (V, E)
\label{simple_graph_definition}
\end{equation}
% Where $V$ is a set of nodes and $E$ contains adjacency matrix. Elements of $E$ are either 1 or 0 to show whether an edge exists between two nodes or not.
We extend the AST for our new representation by including new edges like \texttt{NextToken}, \texttt{NextSib}, and other types that were explained earlier.
We formally define $ParaGraph$ as follows:
\begin{equation}
    ParaGraph = (V, E, T, W)
    \label{paragraph_definition}
\end{equation}
Where $V$ and $E$ are previously defined in Equation~\ref{simple_graph_definition} and $T \in \mathbb{Z}^+$ represents the type of the edges (such as \texttt{NextToken}, \texttt{NextSib} etc).
$W \in \mathbb{Z}^+$ presents the weight which is zero for any edge type other than \texttt{Child}.

These additional edges and attributes are all added to the graphs statically.
Therefore one obvious question could be why we do not let the model reason about these additional attributes by itself rather than presenting them in the graph?
On one hand, adding these attributes to the graphs is not costly; on the other hand, we do not want to waste the resources and the learning capabilities of the model to realize these obvious facts. 
Therefore by augmenting the ASTs and including the attributes mentioned earlier, we can train GNN models more effectively. Later on, our ablation study further improves that the model benefits from the attributes and information presented in $ParaGraph$.
% The GNN model learns more effectively as a result of the addition of information on the graph's connections, types and weights.

We adapt Relational Graph Attention Networks (RGAT) \cite{busbridge2019relational} and use $ParaGraph$ representation as input to train a model for predicting runtime of applications on different accelerators. 
In RGAT, attention logits are computed per each edge type.
In the result section, we will see that with $ParaGraph$, the model has a very small error in its predictions and outperforms the current state-of-the-art approach.

Figure \ref{fig:flow} shows the overall workflow of our GNN-based pipeline for runtime prediction.
The first step is preparing different variants of an application for each accelerator used in this study. This step is further explained in detail in the next section.
Then, using \texttt{Clang}, ASTs are produced and a series of augmentations, as discussed, are applied to them to construct $ParaGraph$.
In order to train a model to predict the runtime of an application, we need to create a dataset with a list of applications and their variant accompanied by their runtime. Therefore, we execute each one of the variants on the specified accelerator and measure the runtime.
Lastly, the $ParaGraph$ representation of variants and their runtime are used to train the GNN model. 
% \textcolor{red}{Apart from the $ParaGraph$ representation, we also let the GNN model know how many teams and threads will be used for the execution of an application.} 
% \textcolor{blue}{
Along with $ParaGraph$s, our feature set also includes the number of teams and threads used for executing an application.
These features are then fed into the GNN-based model for predicting the runtime.
% }
% \textcolor{red}{Therefore, GNN model makes predictions based on $ParaGraph$ representation of the applications and the number of threads and teams.}

% https://docs.google.com/drawings/d/1jTC8KaCykPDVlHKipV1gyWBfPcxOEdEzaRGJWqbKMd4/edit?usp=share_link

% \subsection{Examples}
 % 2pg
\section{Experiments and Setups}
\label{sec:experiment}
We used two clusters and compilers to test and evaluate our tool.
Our experiments were carried out on the ORNL's Summit supercomputing cluster~\cite{Summit} with LLVM/Clang (ver 13.0) and a \texttt{nvptx} backend for GPUs, as well as the LLNL's Corona cluster~\cite{Corona} with LLVM/Clang (ver 15.0) and a \texttt{rocm} backend for GPUs.
For the purposes of this study, we only use one GPU per cluster node.
%Future research will address the management of multiple GPUs.

% The biggest obstacle any Machine Learning engineer working on compiler problems faces is the lack of publicly accessible data, and we were no exception to that.
One of the primary obstacles we faced, as with any other data-driven approach, is the lack of a publicly accessible dataset.
Creating a dataset that can be used to train our model was the first step in building a data-driven based cost model.
Although there are plenty of OpenMP benchmark suites publicly available, little effort has been put into OpenMP GPU offloading benchmarking.
There are very few publicly available benchmarks for OpenMP GPU offloading.
Even though a few benchmarks, such as the \texttt{Rodinia} benchmark suite~\cite{che2009rodinia}, include kernels that implement OpenMP GPU offloading, we cannot use them due to a lack of different kernel variations.
Therefore, as a first step, we create a collection of benchmarks that leverage OpenMP GPU offloading.
The goal is to include a broad class of benchmarks that would cover a spectrum of statistical simulation, linear algebra, data mining, etc.
% For this we developed application tabulated in Table~\ref{tab:benchmark}.

\subsection{Data Collection}

There are three parts to our data collection: Code Variant Generation, Graph Generation, and Runtime Collection.

\begin{table}[ht]
\centering
\begin{tabular}{|L{0.45\columnwidth}|C{0.15\columnwidth}|L{0.25\columnwidth}|}
\hline
\rowcolor{black}
\textcolor{white}{\textbf{Application}} & \textcolor{white}{\textbf{Num Kernels}} & \textcolor{white}{\textbf{Domain}} \\ \hline
\textbf{Correlation Coefficient} ~\cite{schober2018correlation} & 1 & Statistics \\ \hline

\textbf{Covariance} & 2 & Probability Theory \\ \hline

\textbf{Gauss Seidel Method} & 1 & Linear Algebra  \\ \hline

\textbf{K-nearest neighbors}~\cite{che2009rodinia} & 1 & Data Mining \\ \hline

\textbf{Laplace's Equation}~\cite{mitra2010finite} & 2 & Numerical Analysis \\ \hline

\textbf{Matrix-Matrix Multiplication} & 1 & Linear Algebra \\ \hline

\textbf{Matrix-Vector Multiplication} & 1 & Linear Algebra \\ \hline

\textbf{Matrix Transpose} & 1 & Linear Algebra \\ \hline

\textbf{Particle Filter}~\cite{che2009rodinia} & 7 & Medical Imaging  \\ \hline
\end{tabular}
\caption{Benchmark Applications} 
\label{tab:benchmark}
\vspace{-2em}
\end{table}
\subsubsection{Code Variant Generation}
The first step is to create different kernels for which data needs to be collected.
We needed application kernels from multiple domains to have a better prediction across diverse disciplines of Computer Science.
% These applications are already defined in Section~\ref{subsec:benchmark}.
Table~\ref{tab:benchmark} specifies nine benchmark applications that we have identified for data collection.
However, nine applications are not enough.
There are only seventeen kernels if we count the number of kernels in each of these applications.
In the future, as this model develops, more applications will be added.

As a proof of concept for this research, we only consider the following six transformations:
\begin{itemize}[nosep,leftmargin=*]
    \item \textbf{cpu}: A cpu parallel kernel using \texttt{omp~parallel~for}.
    \item \textbf{cpu\_collapse}: In case of a nested collapsible loop, we collapse it with \texttt{omp parallel~for~collapse(2)} directive.
    \item \textbf{gpu}: A gpu kernel using a combined \texttt{omp target teams distribute parallel for} directive. All data is already considered to be present on the GPU.
    \item \textbf{gpu\_collapse}: A gpu kernel with nested collapsible loop using \texttt{omp target teams distribute parallel for collapse(2)} directive. All data is already considered to be present on the GPU.
    \item \textbf{gpu\_mem}: Same as combined gpu offloading, but with data transfer.
    \item \textbf{gpu\_collapse\_mem}: Same as combined gpu\_collapse, but with data transfer.
\end{itemize}
We used the OpenMP~Advisor tool~\cite{mishra2023openmp} to generate these kernel variants.
Even after creating multiple variations for each kernel, we only had 66 distinct kernels.
To create more kernels, we vary the levels of parallelism and data used for all these kernels.
We modified these parameters for each application to generate a minimum of 2000-3000 distinct kernels for our dataset.
When the kernels from all applications are combined, we get almost 26,000 unique pieces of information for our dataset.
This step is independent of the architecture on which we will be running the kernels.

\subsubsection{ParaGraph Generation}
In this step of our pipeline, we create the $ParaGraphs$ for the kernels generated in the previous section.
For each kernel that we want to train, our dataset should include two parts -- the GNN graph and the runtime.
For training, the first step is to generate $ParaGraph$ for each one of the kernels.
Since $ParaGraph$ is based on AST, each kernel is compiled, and ASTs are generated.
In order to produce $ParaGraph$, each AST is traversed, and additional edges and weights are added as explained in Section %\ref{paragraph-explained}.
\ref{sec:paragraph}.
% It is important to mention that additional edges with types and also edge weights play an important role in presenting the kernels.
While edge type conveys the information from the compiler's point of view, edge weights expose information about the flow of execution and, more importantly, the hotspots of the kernels, such as loops. 
Typically, edges within a \texttt{ForLoop} have more weights to represent the number of iterations of a loop.

% \begin{table*}[t]
%   \centering
%   \begin{tabular}{|c|c|c|c|}
%     \hline
%     \textbf{Accelerator} & \textbf{\# Data Points} & \textbf{Runtime Range (ms)} & \textbf{Standard Dev} \\
%     \hline
%     NVIDIA V100 GPU & 26040 & [0.035 - 30174] & 3708 \\
%     AMD MI50 GPU & 26668 & [0.448 - 46913] & 4828 \\
%     IBM POWER9 CPU & 13023 & [0.23 - 736798] & 48502 \\
%     AMD EPYC 7401 CPU & 17681 & [0.024 - 291627] & 16942 \\
%     \hline
%   \end{tabular}
%   \caption{Statistics of data points collected on each accelerator.}
%   \label{tab:data_points_stats}
%   \vspace{-2em}
% \end{table*}

\subsubsection{Runtime Collection}
As our next step, we collect the actual runtime of each kernel.
This process is entirely dependent on the hardware that the kernel is intended to run on.%, and it is likely one of the most time-consuming steps of our experiments.
The complexity of this process was determined by our opportunity to get access to the compute nodes of the HPC clusters.
Failures in this step were difficult to handle interactively, and we had to wait for individual jobs to fail before even realizing the error.
As a result, we had to be very cautious when submitting jobs to run on these clusters.
% Before submitting a job, we ensured that all kernels were compilable and that they were running on similar architecture on our local system.
% We used the Seawulf cluster at Stony Brook University as a testbed for all our experiments.
% Seawulf hosts NVIDIA K80 GPU's.
% This GPU is architecturally markedly different from the NVIDIA V100 GPU of Summit or the AMD Radeon Instinct MI50 GPU of Corona, but the process of building and running a kernel on all three GPUs is similar.
% We just need to change some compile-time parameters while building.
% Although the NVIDIA K80 GPUs are much slower than the other GPUs under consideration, we can be confident that if a kernel runs successfully on this GPU, it will also run on the others.

\begin{table}[b]
  \centering
  \vspace{-1em}
  \small
  \begin{tabular}   {|L{0.35\columnwidth}|C{0.1\columnwidth}|C{0.25\columnwidth}|R{0.1\columnwidth}|}
    \hline
    \rowcolor{black}
    \multicolumn{1}{|c|}{\textcolor{white}{\textbf{Platform}}} & \textcolor{white}{\textbf{\#Data Points}} & \textcolor{white}{\textbf{Runtime Range (ms)}} & \multicolumn{1}{|C{0.1\columnwidth}|}{\textcolor{white}{\textbf{Std. Dev.}}} \\
    \hline
    \textbf{Summit}&&&\\
    IBM POWER9 (CPU) & 13,023 & [0.23 - 736,798] & 48,502 \\
    NVIDIA V100 (GPU) & 26,040 & [0.035 - 30,174] & 3,708 \\
    \hline
    \textbf{Corona}&&&\\
    AMD EPYC7401 (CPU) & 17,681 & [0.024 - 291,627] & 16,942 \\
    AMD MI50 (GPU) & 26,668 & [0.448 - 46,913] & 4,828 \\
    \hline
  \end{tabular}
  \caption{Data points collected on each accelerator.}
  \label{tab:data_points_stats}
\end{table}

% \newcolumntype{C}[1]{>{\centering\let\newline\\\arraybackslash\hspace{0pt}}m{#1}}

% \begin{table}[t]
%   \centering
%   \small
%   \begin{tabular}{|l|C{0.12\columnwidth}|C{0.3\columnwidth}|c|}
%     \hline
%     \multicolumn{1}{|c|}{\textbf{Architecture}} & \textbf{\# Data Points} & \textbf{Runtime Range (ms)} & \textbf{Std Dev} \\
%     \hline
%     \multicolumn{4}{|l|}{\textbf{GPU}s}\\
%     \hline
%     NVIDIA V100 & 26040 & [0.035 - 30174] & 3708 \\
%     AMD MI50 & 26668 & [0.448 - 46913] & 4828 \\
    
%     \hline
%     \multicolumn{4}{|l|}{\textbf{CPU}s}\\
%     \hline
%     IBM POWER9 & 13023 & [0.23 - 736798] & 48502 \\
%     AMD EPYC 7401 & 17681 & [0.024 - 291627] & 16942 \\
%     \hline
%   \end{tabular}
%   % \caption{Statistics of data points collected on each accelerator.}
%   \caption{Data points collected on each accelerator.}
%   \label{tab:data_points_stats}
%   \vspace{-2em}
% \end{table}
Using the OpenMP Advisor, we were able to generate code that already collected kernel runtime information.
It accomplished this by including the \texttt{gettimeofday} function call, which gets the clock time of a system expressed in elapsed seconds and microseconds, both before and after a kernel call, and then computes the difference between the times to obtain the runtime in microseconds.
OpenMP Advisor also aided us in generating all six variants of each kernel, and we were able to generate multiple variants for each application by varying the levels of parallelism and data used.
We built each of these kernels on a local cluster to ensure that our code does not break as a result of using an automated tool to generate our kernels.
After successfully building all of our kernels, we ran each kernel on a local cluster to validate our data collection.
Once we were confident that all of our kernels were building and running on a local cluster, we created jobs to build and run them on the Summit and Corona clusters.
Nonetheless, we had limited access to these clusters, and our job would not run for long due to node failure or time constraints.% running out.
~As a result, we had to submit multiple jobs and keep track of their success or failure in collecting data.

As soon as we had each kernel's runtime, we could easily accompany them with the corresponding $ParaGraphs$ to collect data points for our dataset.
Table \ref{tab:data_points_stats} shows the statistics of data points collected from each one of the accelerators.

\subsection{ParaGraph Model}
Once we have enough data on both the Summit and Corona clusters, we begin training our model.
We implemented a GNN-based neural network using RGAT \cite{busbridge2019relational} as the convolution layers. 
Our model is implemented using Pytorch-Geometric library with Mean Squared Error as the loss function, and Adam \cite{kingma2014adam} as the optimizer. % with a learning rate of $10^{-3}$.
% We incorporate a dynamic learning rate that decreases the learning rate with a factor of $0.8$ and patience of $2$ on the RMSE of the validation set.
To embed the graph, the model uses three graph convolution layers based on RGAT, followed by two fully connected layers with \texttt{ReLu} activation function.
As mentioned, the number of teams and threads are considered as two additional features. Another fully connected layer is used to embed these two features.
Finally, the embedding of the graph and the two features are concatenated together and passed through the last fully connected layer for runtime prediction.

The edge weights and two additional features are normalized using MinMaxScaler.
The dataset is split into train-validation sets using a 9:1 ratio. %1pg
\section{Results}
\label{sec:result}
% In this section, experimental results are provided. 
% We compare our approach with \texttt{COMPOFF} \cite{mishra2022compoff} with regard to predicting runtime for GPUs. 
% Unlike COMPOFF, ParaGraph has the advantage of being hardware independent, and has thus been validated on a variety of hardware accelarators including CPUs and GPUs.
In the following subsection, we explain the metrics we have used to evaluate the $ParaGraph$ model.

\subsection{Evaluation Metric}

To evaluate the performance of $ParaGraph$, we use $RMSE$ which is Root Mean Square Error (Equation \ref{rmse_equation}).

\begin{equation}
\label{rmse_equation}
    RMSE = \sqrt{\frac{\sum_{i=1}^{N}(x_i - \hat{x}_i)^2}{N}}
\end{equation}

Where $x_i$ stands for the runtime of a data point in microseconds, $\hat{x}_i$ is the predicted runtime by $ParaGraph$, and $N$ is the total number of samples.

Since the range of runtime differs across platforms, the normalized version of $RMSE$ is also considered. 
Normalized $RMSE$ is calculated by dividing the $RMSE$ by the distance between the minimum and maximum runtime. 
We also use relative error (i.e., absolute error divided by the range of runtime) to report the error rate.

% \begin{table}[b!]
% \centering
% \begin{tabular}{|c|c|c|}
%     \rowcolor{black}
%     \hline
%     \textcolor{white}{\textbf{Accelerator}} & \textcolor{white}{\textbf{RMSE (ms)}} & \textcolor{white}{\textbf{Norm-RMSE}} \\
%     \hline
%     V100 & $280$ & $9\times10^{-3}$ \\
%     MI50 & $510$ & $1\times10^{-2}$ \\
%     POWER9 & $4325$ & $6\times10^{-3}$ \\
%     EPYC7401 & $968$ & $4\times10^{-3}$ \\
%     \hline
% \end{tabular}
% \caption{Experimental Results}
% \label{experimental_results}
% \end{table}

\begin{table}[t!]
\centering
\begin{tabular}{|L{0.4\columnwidth}|c|c|}
    \hline
    \rowcolor{black}
    \multicolumn{1}{|c|}{\textcolor{white}{\textbf{Platform}}} & \textcolor{white}{\textbf{RMSE (ms)}} & \textcolor{white}{\textbf{Norm-RMSE}} \\
    \hline
    \textbf{Summit} &&\\
    IBM POWER9 (CPU) & $4325$ & $6\times10^{-3}$ \\
    NVIDIA V100 (GPU) & $280$ & $9\times10^{-3}$ \\
    \hline
    \textbf{Corona} &&\\
    AMD EPYC7401 (CPU) & $968$ & $4\times10^{-3}$ \\
    AMD MI50 (GPU) & $510$ & $1\times10^{-2}$ \\
    \hline
\end{tabular}
\caption{Experimental Results}
\label{experimental_results}
\vspace{-1.5em}
\end{table}

\begin{figure}[b!]
    \vspace{-1em}
    \includegraphics[width=\columnwidth]{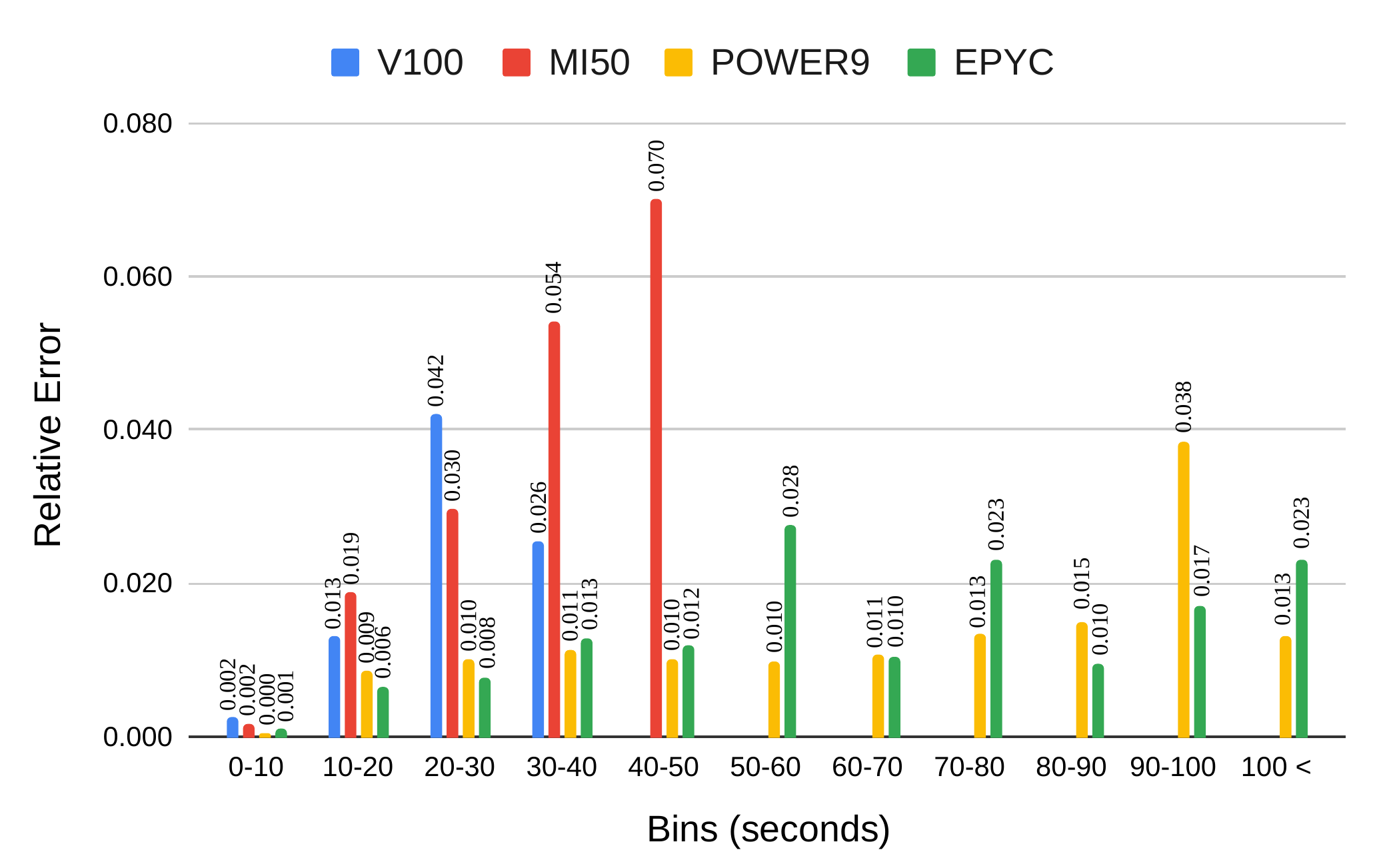}
	\caption{Prediction error per 10-second bins.}
	\label{fig:error_per_bins}
\end{figure}

\subsection{Experimental Results}
\label{sec:exp_results}
Table \ref{experimental_results} shows the experimental results for each accelerator. %They are
We have NVIDIA V100, AMD MI50, and IBM POWER9 with 22 cores and AMD EPYC 7401 with 24 cores.
As shown, the $RMSE$ values range from 280 (ms) to 4325 (ms). The reason why we have different $RMSE$ values, such as 4325 (ms) for POWER9 accelerator, lies in the fact that the runtime dispersion differs across the accelerators. Standard Deviation in Table \ref{tab:data_points_stats} gives us some insights on how dispersed the collected data are. 
Moving on to Normalized-RMSE, which is independent of the range of runtime, we see $ParaGraph$ has relatively the same error across accelerators; thus, it can be applied to different accelerators.

To further analyze our results, we have calculated a relative error per bins of 10 seconds. Figure \ref{fig:error_per_bins} shows 11 bins for all four accelerators. Each bin has a 10-second range except the last bin.
The figure shows that the relative error is small across different bins and accelerators (less than \%10), meaning that our $ParaGraph$ model has stable behavior for different ranges of runtimes.

\begin{figure}[t!]
    \includegraphics[width=\columnwidth]{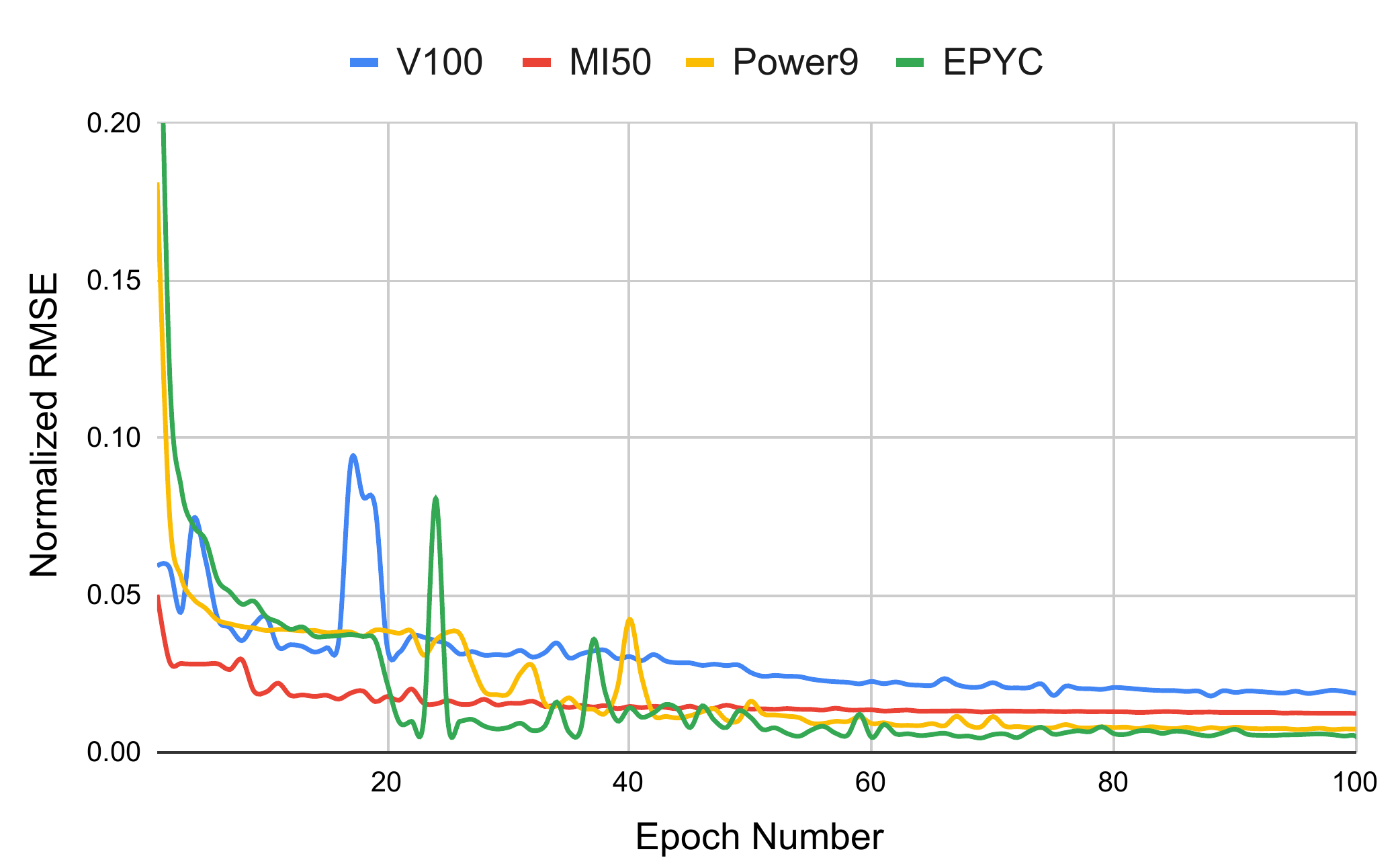}
	\caption{Normalized RMSE per each epoch.}
	\label{fig:validation_per_epoch}
    \vspace{-1em}
\end{figure}

Moreover, we analyze how stable the model is during the training process in Figure \ref{fig:validation_per_epoch}. The figure shows validation $RMSE$ for four accelerators. In the first few epochs, the $ParaGraph$ model is not very stable, resulting in fluctuations in the $RMSE$; however, as the model is trained further, it is able to pick and learn the features from the representation and reduce $RMSE$ value per each epoch and ultimately converges. 

% \textcolor{blue}{To further analyze our results, we have used equal frequency binning to divide the validation set into $10$ bins in increasing order of runtime (bin 0 has the smallest runtimes, while bin 9 has the largest).
% We assume that larger runtimes correspond to larger problems of increasing complexity.
% As shown in Figure \ref{fig:rmse_bins}, our approach predicts runtimes for problems of most complexities quite well.
% The y-axis of the chart in Figure \ref{fig:rmse_bins} uses logarithmic scale for better presentation.
% For reference, the gray bar in bin 9 for CPU (tallest bar) has an RMSE error of $\sim9000$ms for a set of applications and inputs with geometric mean runtimes of $\sim60$ seconds.}

% As can be seen, training the model on the data points collected on AMD GPU results in the lowest $RMSE$.
% Nvidia V100 comes second, and lastly, CPU has the higher $RMSE$ among the others.
% The results indicate that AMD has a more stable runtime with relatively similar behavior across various executions of applications.
% During training our model, we observed that it is capable of learning the characteristics of the applications by minimizing $MSE$ per each epoch.

% \begin{figure}[t!]
%     \includegraphics[width=\columnwidth]{figures/results/rmse_bins.pdf}
% 	\caption{\textcolor{blue}{RMSE in ascending order of problem size/complexity (left to right). Figure explained in Section \ref{sec:exp_results}.}}
% 	\label{fig:rmse_bins}
% \end{figure}

\begin{figure}[b!]
    \vspace{-0.5em}
    \includegraphics[width=\columnwidth]{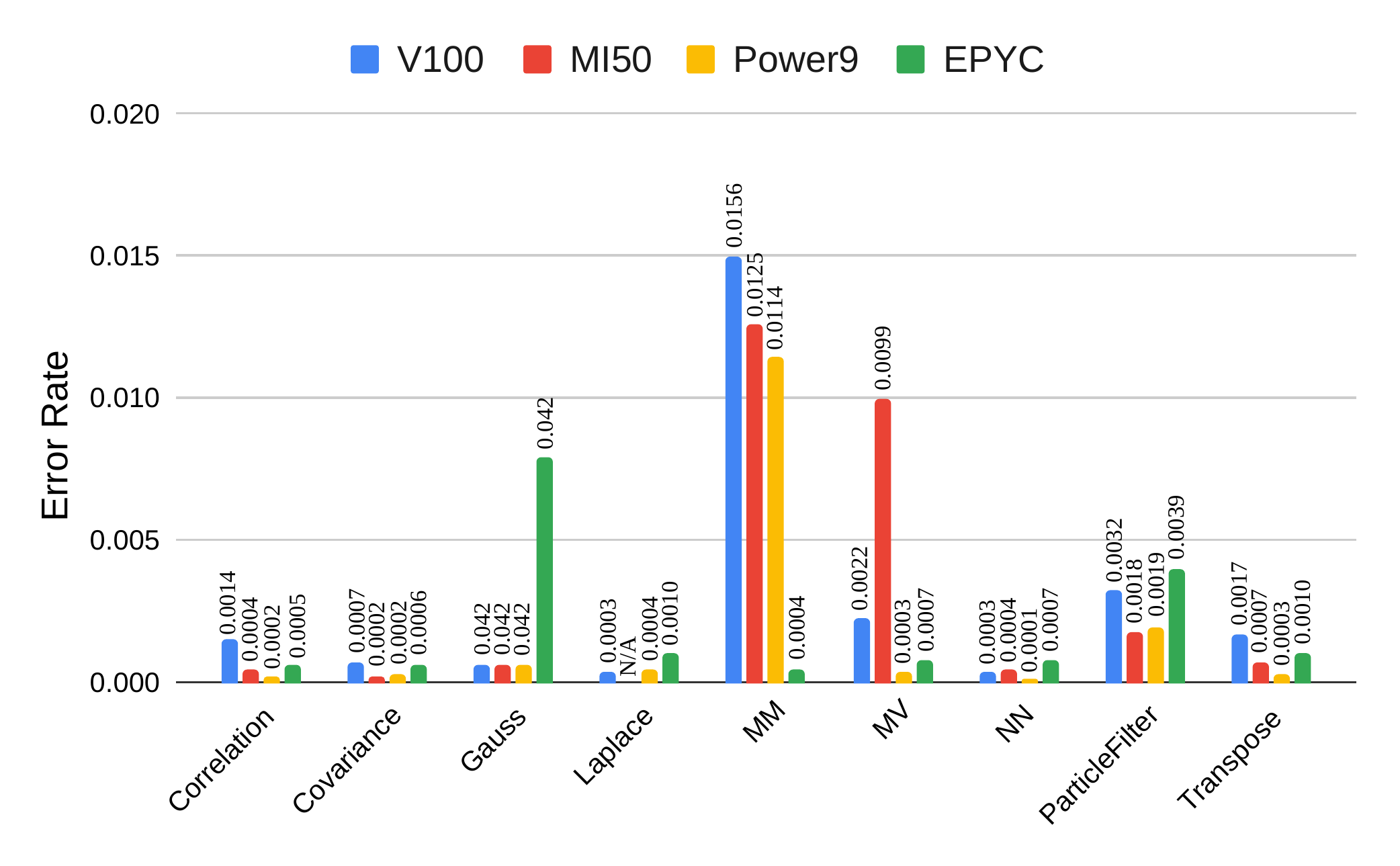}
	\caption{Error rate per each application.}
	\label{fig:error_rate_per_application}
\end{figure}

% \begin{figure}[t!]
%     \includegraphics[width=\columnwidth]{figures/results/Training_data_distribution.pdf}
% 	\caption{Runtime Distribution of Training Data}
% 	\label{fig:training_distribution}
% \end{figure}

% Moving on to Normalized-RMSE, we see the data points on CPU achieved the lowest error. 
% This shows that the reason for the high RMSE on CPU data points is due to the diversity of the measured runtime.
% In Table \ref{tab:data_points_stats}, we can see that the range of runtime for CPU data points is quite large.
% We further analyzed the distribution of data points collected from CPU execution.
% In fact, by looking at Figure \ref{fig:training_distribution} we see that for Nvidia and AMD the majority of data points have a runtime of less than 10 seconds, whereas, for CPU, data points have a wide range of runtime, making the task of prediction more challenging for the model.

% However, our model can still automatically pick and learn the features of the applications via the proposed representation. 
% Figure \ref{fig:validation_per_epoch} shows Normalized $RMSE$ for each epoch for the three accelerators. 
% As expected, for data points from CPU, the model has difficulties lowering the $RMSE$. 
% Despite that, we can see there is a downward trend showing. 
% This shows that the model is indeed able to learn from the \texttt{ParaGraph} representation.

Lastly, we calculate the average relative error per application to see if the $ParaGraph$ model is able to have prediction with low error for all types of applications. Figure \ref{fig:error_rate_per_application} shows the error rate of each application. As can be seen, the model is indeed capable of having accurate predictions for a wide variety of applications resulting in a low error rate. Therefore, the model is not biased toward one application.
On the AMD MI50 GPU, the Laplace data was corrupted during collection.
Consequently, neither this study nor the training process includes that data. 
%We use the CPU data because they are accurate.

% Lastly, Figure \ref{fig:error_rate_per_application} shows the average relative error per each application. Overall, the average error rate is very low (less than 2\%).
% As shown, for some applications such as Correlation, Covariance, Gauss, NN, the average error rate is close to zero.
% On the other hand, ParticleFilter and MM have a slightly higher error rate than other applications (still very low error rates overall). 

% One way of mitigating this challenge is collecting more data points on CPU, which we have left as a future work.

% \subsection{Cross-Architecture Transfer Learning}
% \textcolor{red}{TO REMOVE}

% \begin{figure}[t!]
%     \includegraphics[width=\columnwidth]{figures/results/Runtime Prediction (V100).pdf}
% 	\caption{Runtime prediction}
% 	\label{fig:flow}
% \end{figure}

% \begin{figure}[t!]
%     \includegraphics[width=\columnwidth]{figures/results/Relative Error Per Application (V100).pdf}
% 	\caption{Relative error per each application}
% 	\label{fig:flow}
% \end{figure}

\subsection{Ablation Study}
$ParaGraph$ representation, as explained in Section~\ref{sec:paragraph}, is built by applying a series of major augmentations on top of AST. 
In this section, we quantify the impact of these augmentations.
%We compare $ParaGraph$ representation against two other representions, which are Raw AST and Augmented AST. 
First, we consider the AST itself without additional edges and weights, we call it Raw AST.
Then, we add additional edges and edge types to the AST and name it Augmented AST. 
Lastly, we have $ParaGraph$ that contains both additional edges and also edge weights.
% We consider Raw AST, which is basically a simple AST with only one type of edge, $child$ edge.
% One the other hand, we have Augmented AST is Raw AST with seven other types of edges that were described in Section \ref{sec:augment_ast} and Figure \ref{augmentedAST}.
Table \ref{tab:ablation_study} shows the results of the ablation study.
We see that Raw AST results in the highest error for all four accelerators.
Adding new edges and introducing new types of edges (Augmented AST) improve the prediction to some extent.
For example, the $RMSE$ of V100 drops from 2114 (ms) to 786 (ms) with the addition of these new edges.

One of the key characteristics of our proposed program representation is the edge weights. 
Edge weights convey essential information about how often different regions of the AST execute.
Therefore, we see quite a good improvement in $RMSE$ when edge weights are added. For instance, the $RMSE$ for V100 is further improved to 280 (ms).

\begin{table}[h]
  \centering
  \small
  \begin{tabular}   {|L{0.35\columnwidth}|C{0.13\columnwidth}|C{0.13\columnwidth}|R{0.19\columnwidth}|}
    \hline
    \rowcolor{black}
    \multicolumn{1}{|c|}{\textcolor{white}{\textbf{Platform}}} & \textcolor{white}{\textbf{Raw AST}} & \textcolor{white}{\textbf{Aug AST}} & \multicolumn{1}{|C{0.19\columnwidth}|}{\textcolor{white}{\textbf{ParaGraph}}} \\
    \hline
    \textbf{Summit}&&&\\
    IBM POWER9 (CPU) & $27593$ & $26860$ & $\textbf{4325}$ \\
    NVIDIA V100 (GPU) & 2114 & 786 & \textbf{280} \\
    \hline
    \textbf{Corona}&&&\\
    AMD EPYC7401 (CPU) & $11911$ & $9633$ & $\textbf{968}$ \\
    AMD MI50 (GPU) & $2888$ & $1177$ & $\textbf{510}$ \\
    \hline
  \end{tabular}
  \caption{RMSE of training with and without edges' weight.}
  \label{tab:ablation_study}
  \vspace{-2em}  
\end{table}

% \begin{table}[h]
% \centering
% \begin{tabular}{c|c|c|c}
%     \hline
%     \textbf{Accelerator} & \textbf{Raw AST} & \textbf{Augmented AST} & \textbf{ParaGraph}\\
%     \hline
%     V100 & $2114$ & $786$ & $\textbf{280}$ \\
%     MI50 & $2888$ & $1177$ & $\textbf{510}$ \\
%     Power9 & $27593$ & $26860$ & $\textbf{4325}$ \\
%     EPYC & $11911$ & $9633$ & $\textbf{968}$ \\
%     \hline
% \end{tabular}
% \caption{RMSE of training with and without edges' weight. Removing the edge weights from graphs increases the error rate.}
% \label{tab:ablation_study}
% \end{table}

\begin{figure}[b!]
    \includegraphics[width=\columnwidth]{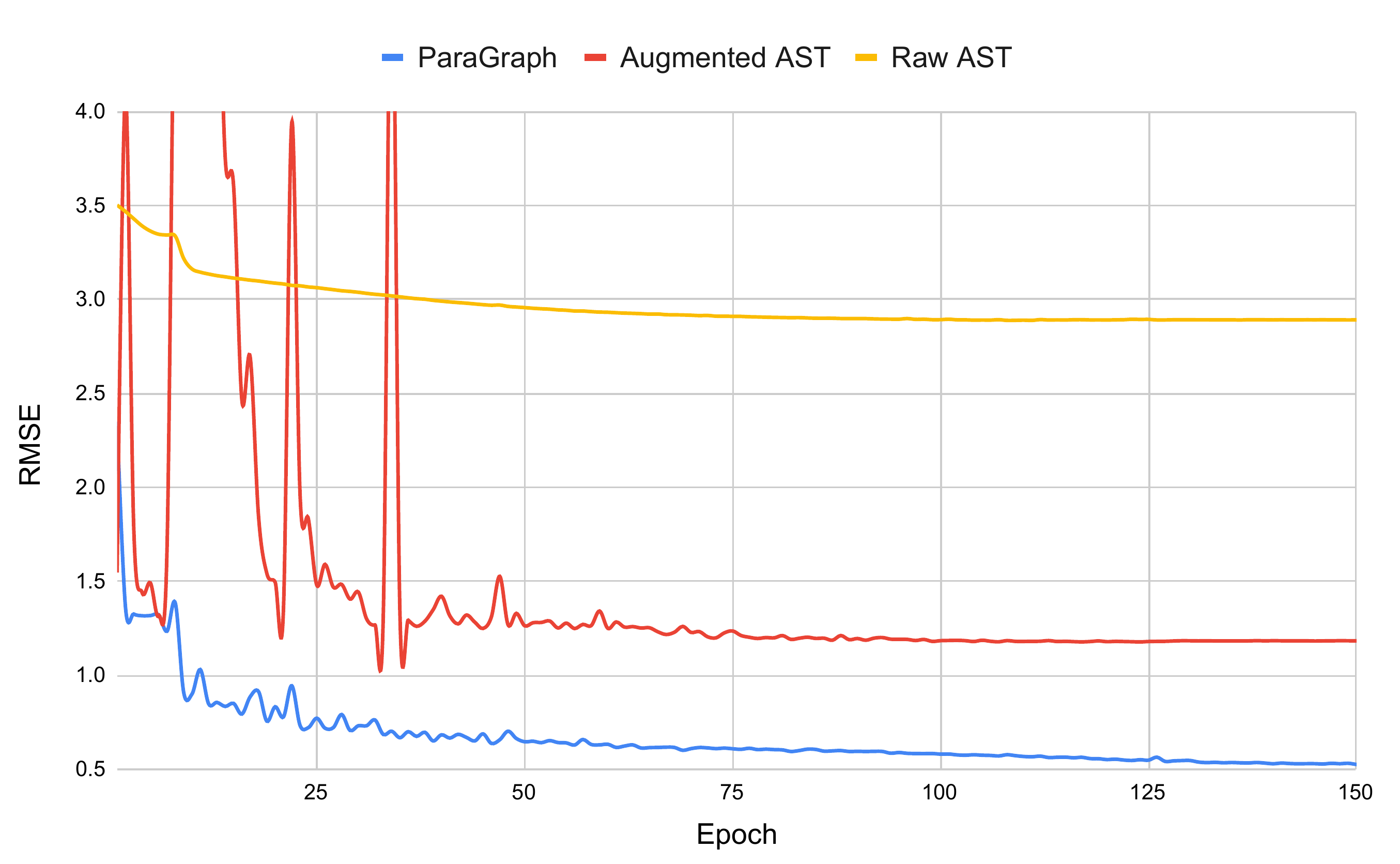}
	\caption{RMSE of the validation set during training the GNN models on MI50 data points.}
	\label{fig:ablation_training_process}
    \vspace{-1em}
\end{figure}

We analyze the addition of edges and their weights further to see how the training process of the $ParaGraph$ model is affected by these augmentations.

Figure \ref{fig:ablation_training_process} shows how the model behaves during training. This figure depicts the $RMSE$ value per each epoch for Raw AST, Augmented AST, and $ParaGraph$ on the MI50 accelerator. 

Using only the Raw AST without any augmentations, which means having only one edge type, the model is able to learn some characteristics of the applications and reduce the $RMSE$ per epoch; however, this reduction in $RMSE$ is not significant. 
Augmented AST contains 8 different types of edges. We see the addition of these edges destabilized the training process of the model. In the first few epochs, the model is challenged to learn different relations between the nodes.
However, eventually, after several epochs, the prediction of the model is stabilized and it achieves $RMSE$ of 1177 (ms).
Once edges of the Augmented AST are augmented with weights, thus $ParaGraph$ is constructed, we see further improvements in the model's predictions. Although the validation $RMSE$ has some fluctuation in the initial epochs, it ultimately converges with a considerably smaller error.

% \begin{table}[h]
% \centering
% \begin{tabular}{c|c|c|c}
%     \hline
%     \textbf{Accelerator} & \textbf{Raw AST} & \textbf{Augmented AST} & \textbf{ParaGraph}\\
%     \hline
%     V100 & $2114$ & $786$ & $\textbf{280}$ \\
%     MI50 & $2888$ & $1177$ & $\textbf{510}$ \\
%     Power9 & $27593$ & $26860$ & $\textbf{4325}$ \\
%     EPYC & $11911$ & $9633$ & $\textbf{968}$ \\
%     \hline
% \end{tabular}
% \caption{RMSE of training with and without edges' weight. Removing the edge weights from graphs increases the error rate.}
% \label{tab:ablation_study}
% \end{table}

% Create a plot with Bins and the Error Rate per BINs.

% Table \ref{tab:ablation_study} shows the RMSE value of training the models with and without edge weights.
% We can see that edge weights are one of the important features in our program representation, without them, our model will not be able to have accurate predictions.

\subsection{Comparison with State-of-the-art Tool}
% Make this more general, 'comparison with the state of the approach'
To the best of our knowledge, $COMPOFF$~\cite{mishra2022compoff} is the only state-of-the-art OpenMP GPU offloading cost model.
We further compare the results from $ParaGraph$ with those of $COMPOFF$.
As mentioned in Section~\ref{sec:ompadvisor} OpenMP Advisor uses $COMPOFF$ for predicting runtime for OpenMP kernels.

While OpenMP Advisor eventually needs a cost model that can forecast for all possible underlying architectures, $COMPOFF$ is currently only suitable for GPU execution.
In contrast, $ParaGraph$ can model the application runtime regardless of the underlying architecture.
In our experiments, $ParaGraph$ was used to predict runtime on both CPUs and GPUs.
Therefore, we compare $ParaGraph$ and $COMPOFF$ using data points collected from the NVIDIA V100 GPU only.

\begin{figure}[t!]
    \includegraphics[width=\columnwidth]{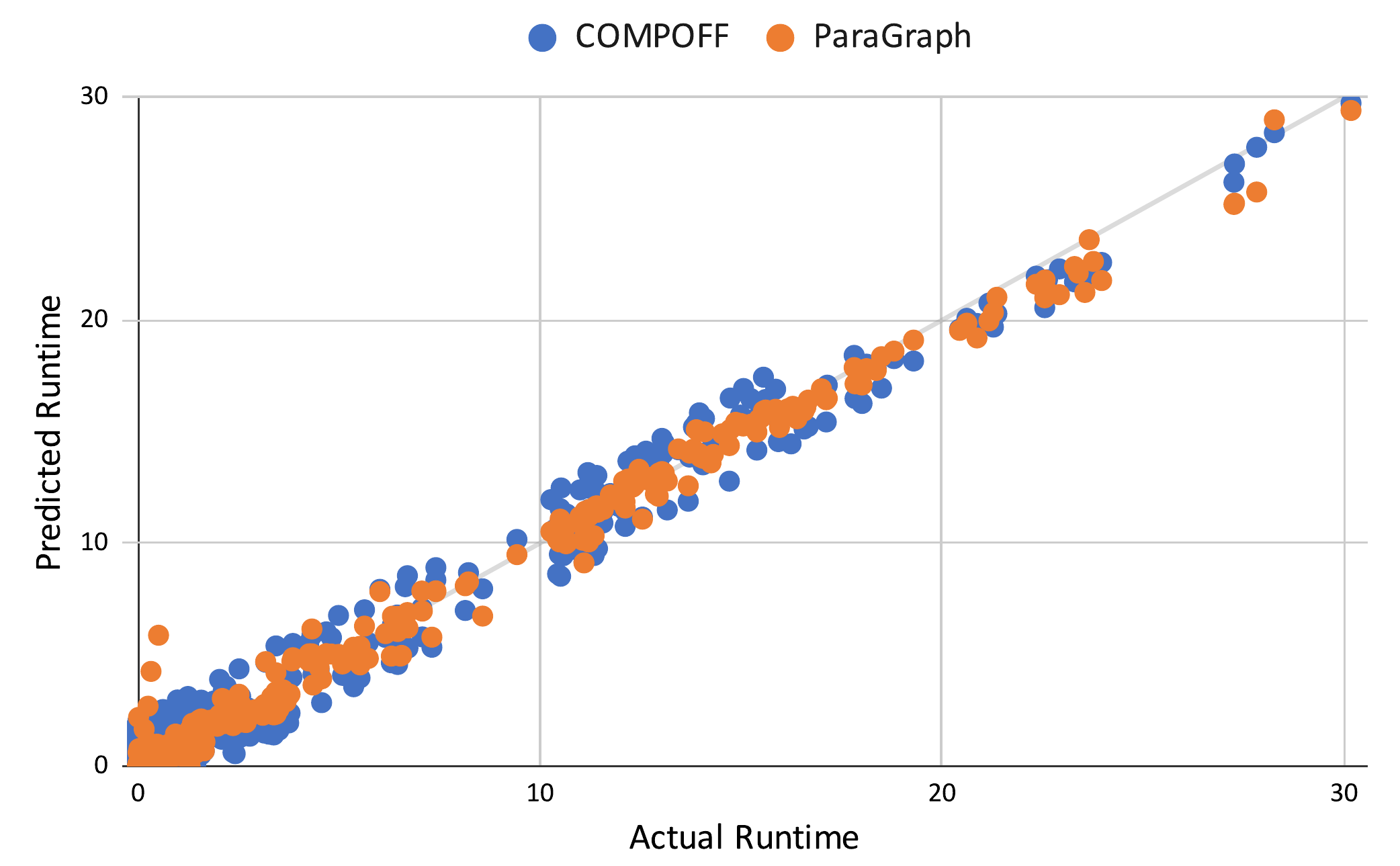}
    \caption{Comparison of ParaGraph and COMPOFF on predicting runtimes on NVIDIA V100 for each data point.}
	\label{fig:ParaGraph_vs_COMPOFF_Actual}
    \vspace{-1.5em}
\end{figure}

\begin{figure}[t!]
    \includegraphics[width=\columnwidth]{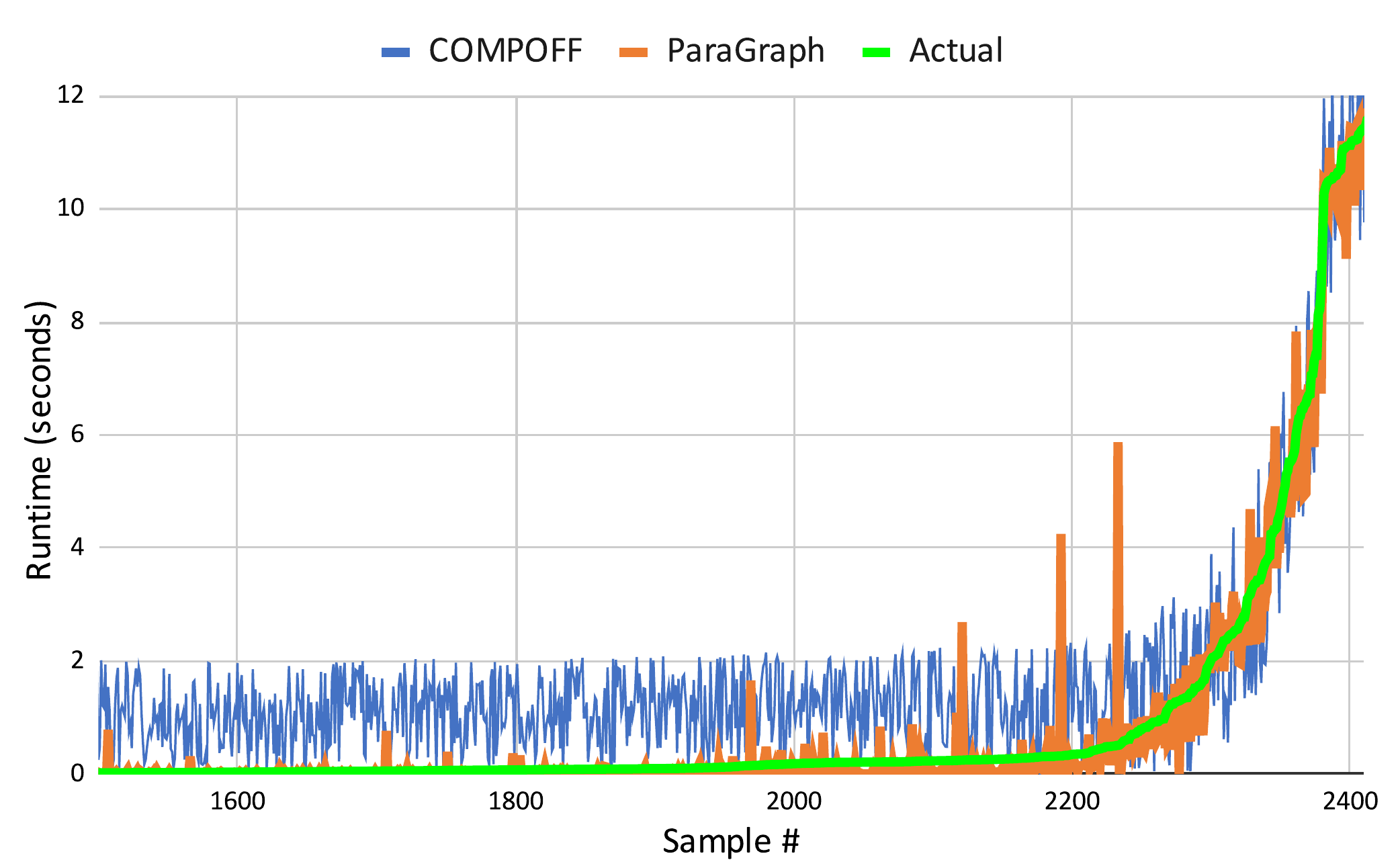}
    \caption{ParaGraph and COMPOFF prediction on NVIDIA V100 as compared to the actual runtime.}
	\label{fig:ParaGraph_vs_COMPOFF}
    \vspace{-1em}
\end{figure}

For both $ParaGraph$ and $COMPOFF$, Figure~\ref{fig:ParaGraph_vs_COMPOFF} demonstrates a strong correlation between the actual and predicted data.
The results for $COMPOFF$ are represented by blue dots, while those for $ParaGraph$ are represented by orange dots.
Despite the fact that both perform admirably, $ParaGraph$ demonstrates a much stronger correlation between the predicted and actual runtime.
In Figure~\ref{fig:ParaGraph_vs_COMPOFF_Actual}, $COMPOFF$ (blue) demonstrates a slightly higher error rate for smaller runtime kernels, but as the runtime increases, this error rate decreases (or disappears entirely).
However, for all kernels, the error rate is significantly lower for $ParaGraph$ (orange).
Similar results were also observed on the AMD GPU.
We were unable to compare the outcomes of our prediction on CPUs because $COMPOFF$ does not support CPUs.
This is one significant advantage $ParaGraph$ has over $COMPOFF$.

% To further analyze our study results, we divided the validation set into 7 bins of 5 seconds each, in increasing order of runtime.
% The average runtime, average $ParaGraph$ prediction, and average $COMPOFF$ predictions are all plotted together in Figure~\ref{fig:ParaGraph_vs_COMPOFF_runtime_prediction}.
% Here, we can clearly see that $COMPOFF$'s predictions for kernels with runtimes of under 5 seconds are worse.
% However, $ParaGraph$ performs admirably across the entire kernel range.

% \begin{figure}[t!]
%     \begin{subfigure}[t]{0.48\columnwidth}
%     \includegraphics[width=\columnwidth]{figures/results/compoff_paragraph_scatter_chart.pdf}
%     \caption{}
%     \end{subfigure}	
%     \hfill
%     \begin{subfigure}[t]{0.48\columnwidth}
%     \includegraphics[width=\columnwidth]{figures/results/compoff_paragraph_line_chart.pdf}
%     \caption{}
%     \end{subfigure}	
%     \caption{Comparison of ParaGraph and COMPOFF on predicting runtimes on NVIDIA V100 for each data point.}
% 	\label{fig:ParaGraph_vs_COMPOFF}
% \end{figure}

% \begin{figure}[t!]
%     \includegraphics[width=\columnwidth]{figures/results/ParaGraph_vs_COMPOFF_runtime_prediction.pdf}
% 	\caption{Comparison of ParaGraph and COMPOFF on predicting runtimes per each bin.}
% 	\label{fig:ParaGraph_vs_COMPOFF_runtime_prediction}
% \end{figure}

% \begin{figure}[t!]
%     \includegraphics[width=\columnwidth]{figures/results/COMPOFF_Para_Per_app.pdf}
% 	\caption{Relative error of ParaGraph and COMPOFF per applications.}
% 	\label{fig:ParaGraph_vs_COMPOFF_relative}
% \end{figure}

\subsection{Discussion} % Limitation
$ParaGraph$ is able to model the execution of applications statically.
Our experimental results show the effectiveness of $ParaGraph$ in predicting applications' runtime across different accelerators. 
On the other hand, there are applications whose runtime behavior is not stable. 
That is, they show different behaviors across various executions. 
These applications can not be modeled statically. 
As a result, predicting their runtime using only static features is a challenge. 
One way of tackling this challenge could be developing a hybrid approach using static and dynamic features (for example, Performance Counters) of applications.
However, collecting dynamic features of applications requires the execution of applications which will potentially increase the cost of prediction. 

% \begin{figure}[t!]
%     \includegraphics[width=\columnwidth]{figures/results/RMSE_20230119-132246.pdf}
% 	\caption{RMSE with MSE Loss}
% 	\label{fig:flow}
% \end{figure}

% \begin{figure}[t!]
%     \includegraphics[width=\columnwidth]{figures/paragraph_compoff.png}
% 	\caption{Comparison of ParaGraph to COMPOFF}
% 	\label{fig:paragraph_compoff_comparison}
% \end{figure} % 2pg
\section{Conclusion and Future Work}
\label{sec:conclusion}
In this paper, we proposed $ParaGraph$. A novel way of representing HPC kernels. 
$ParaGraph$ is based on ASTs; it incorporates additional features to the AST to better represent applications.
In particular, $ParaGraph$ includes some of the insights of the compiler by adding new edges to the AST.
$ParaGraph$ also adds information about the way loops and if statements will be executed by adding weights to the corresponding edges.
We evaluated $ParaGraph$ on a set of applications for four different accelerators. 
It achieved a low error rate highlighting the effectiveness of $ParaGraph$.
One of the main benefits of $ParaGraph$ is its independence of underlying hardware architecture. Therefore, it can be applied to a wider range of architectures, unlike previous approaches.
%Abid

In this work, we used $ParaGraph$ for predicting the execution time of a given kernel for OpenMP GPU offloading problem. 
We plan to explore and analyze how $ParaGraph$ can help other OpenMP optimization strategies such as predicting SIMD stride, scheduling strategies, loop chunk size, etc. in the future. Another interesting research area to explore is to use $ParaGraph$ and capture parallelism for other parallel programming models such as OpenACC, Kokkos, HIP, etc. for building cost models for various optimization problems.

% Abid
% Future work seems incomplete
 %0.5pg

\bibliographystyle{IEEEtran}
\bibliography{text/bibliography.bib} % 1pg

%Comment the following line to remove the sample text
% \textcolor{blue}{Following is just a sample text.}
% \input{sample_files/sample_text}

\end{document}